\documentclass[namedreferences]{SolarPhysics}
\usepackage[optionalrh,solaenum]{spr-sola-addons} 
\usepackage{graphicx}        
\usepackage{color}           
\usepackage{url}             

\begin{document}

\begin{article}

\begin{opening}

\title{Multiwavelength study of a Solar Eruption from AR NOAA 11112 I. Flux Emergence, Sunspot Rotation and Triggering of a Solar Flare}
\author{Pankaj~\surname{Kumar}$^{1}$\sep
	Sung-Hong~\surname{Park}$^{1}$\sep
        K.-S.~\surname{Cho}$^{2,3,1}$\sep       
        S.-C.~\surname{Bong}$^{1}$
                      	       }
\runningauthor{P. Kumar et al.}
\runningtitle{Flux emergence, Sunspot rotation and Triggering of Solar Flare}
\institute{$^{1}$ Korea Astronomy and Space Science Institute (KASI), Daejeon, 305-348, Republic of Korea.
                     email: \url{pankaj@kasi.re.kr}}
\institute{$^{2}$ NASA Goddard Space Flight Center, Greenbelt, Maryland, USA.}
\institute{$^{3}$ Department of Physics, The Catholic University of America, Washington, D. C., USA.}
\begin{abstract}
 We analyse the multiwavelength observations of an M2.9/1N flare that occurred in the active region (AR) NOAA 11112 in the vicinity of a huge filament system on 16 October 2010. 
SDO/HMI magnetograms reveal the emergence of a bipole (within the existing AR) 50 hours prior to the flare event.  During the emergence, both the positive and negative sunspots in the bipole show translational as well as rotational motion. The positive polarity sunspot shows the significant motion/rotation in the south-westward/clockwise direction and continuously pushing/sliding the surrounding opposite polarity field region. On the other hand, the negative polarity sunspot moves/rotates in the westward/anticlockwise direction. The positive polarity sunspot rotates $\approx$$70^\circ$ within 30 hours, whereas negative polarity $\approx$$20^\circ$ within 10 hours. 
SDO/AIA 94 \AA \ EUV images show the emergence of a flux tube in the corona consistent with the emergence of the bipole in HMI. The footpoints of the flux tube were anchored in the emerged bipole. The initial brightening starts at one of the footpoint (western) of the emerged loop system, where the positive polarity sunspot pushes/slides towards a nearby negative polarity field region. A high speed plasmoid ejection (speed$\approx$1197 km s$^{-1}$) was observed during the flare impulsive phase, which suggests the magnetic reconnection of the emerged positive polarity sunspot with the surrounding opposite polarity field region.  
The entire AR shows the positive helicity injection before the flare event. Moreover, the newly emerging bipole reveal the signature of negative (left-handed) helicity.
These observations provide the unique evidences of the emergence of twisted flux tube from below the photosphere to coronal heights triggering a flare mainly due to the interaction between the emerged positive polarity sunspot and a nearby negative polarity sunspot by the shearing motion of the emerging positive sunspot towards the negative one. Our observations also strongly support the idea that the rotation is most likely attributed to the emergence of twisted magnetic
 fields, as proposed by recent models.
\end{abstract}
\end{opening}

\section{Introduction}
Solar flares are one of the transient phenomena of magnetic energy release in the solar atmosphere. The magnetic energy stored in the twisted/sheared complex magnetic fields of active regions is converted into thermal and kinetic energies, as well as the acceleration of energetic particles via magnetic reconnection \cite{priest02,chen11}. The emergence/activation of twisted flux tubes/ropes can lead to magnetic instabilities, interacting with the overlying fields and eventually result in the flare and associated eruption \cite{cho2009,kumar2010b,srivastava2010,foullon2011,kumar2011}. Using two dimensional magnetohydrodynamic (MHD) numerical simulation, \inlinecite{chen2000} proposed an emerging flux trigger mechanism for the onset of CMEs based on the flux rope model. Kink instability of the twisted magnetic flux ropes may also be responsible for the initiation of some solar eruptions \cite{torok2003,torok2005,kumar2012}, and the interaction/reconnection between filaments/flux ropes may cause the eruption of the twisted magnetic fields \cite{kumar2010a,torok2011}. In addition, the coalescence of magnetic loops caused by footpoint shearing motion may also play an important role in flare triggering \cite{sakai1986,kumar2010c}. However, it is essential to investigate more exact mechanisms for the initiation of solar eruptive phenomena, because each flare and the associated dynamical processes may be unique in the energy build-up and release in associated active regions.

Rotating sunspots may be one of the most likely candidates to inject magnetic helicity into the solar atmosphere, which may cause the flare energy build-up \cite{stenflo1969,barnes1972,amari1996,tokman2002,torok2003,regnier2006,zhang2008}. Many pieces of observational evidence of sunspot rotation have been reported (e.g. several hundred degrees around their umbral centers over a few days) prior to the occurrence of strong flares \cite{brown2003,zhang07,su2008,yan09,min2009,kumar2010a,chandra2011}. The rapid rotation
 of the sunspots  can cause the formation of sigmoids loop that can erupt to produce flares and coronal mass ejections \cite{canfield1999,pevtsov2002,yan2007}. The recent theories/models suggest that sunspot rotation can
 be caused by a pre-twisted magnetic flux tube emerging from below the photosphere \cite{longcope2000,gibson2004,magara2006,fan2009}.
\inlinecite{tian2006} found sunspot rotation as a primary driver of helicity production and injection into the corona and suggested
 that the observed active region dynamics, subsequent filament and sigmoid eruption are driven by
 a kink instability which occurred due to a large amount of the helicity injection.
 \inlinecite{tian2008} suggested that the emergence of significantly twisted magnetic fields may not always result in the rotation of the associated sunspots, but they do play a very important role in the coronal helicity accumulation and free-energy build-up. Recently, \inlinecite{jiang2012} studied the sunspot evolution/rotation associated with the first X-class flare of the current solar cycle, which occurred in AR 11158 on 2011 February 15 and suggested that the shearing and rotational motions are the main
 contributors to the energy build-up and helicity injection leading to the flare, the cancellation and collision might act as a
 trigger. However, the exact cause and effect of rotating sunspots are not well understood up to now.

In this paper, we analyse the multiwavelength observations from {\it Solar Dynamic Observatory} (SDO: \opencite{pesnell2012}) to investigate the possible cause of energy build-up and triggering mechanism of a M-class flare that occurred in the AR NOAA 11112. 
 In Section 2, we present multiwavelength observations of the M-class flare. In Section 3, we describe the evolution of magnetic field. In the last section, we discuss the results and draw the conclusions.  

\begin{figure}
\centerline{
\includegraphics[width=0.9\textwidth]{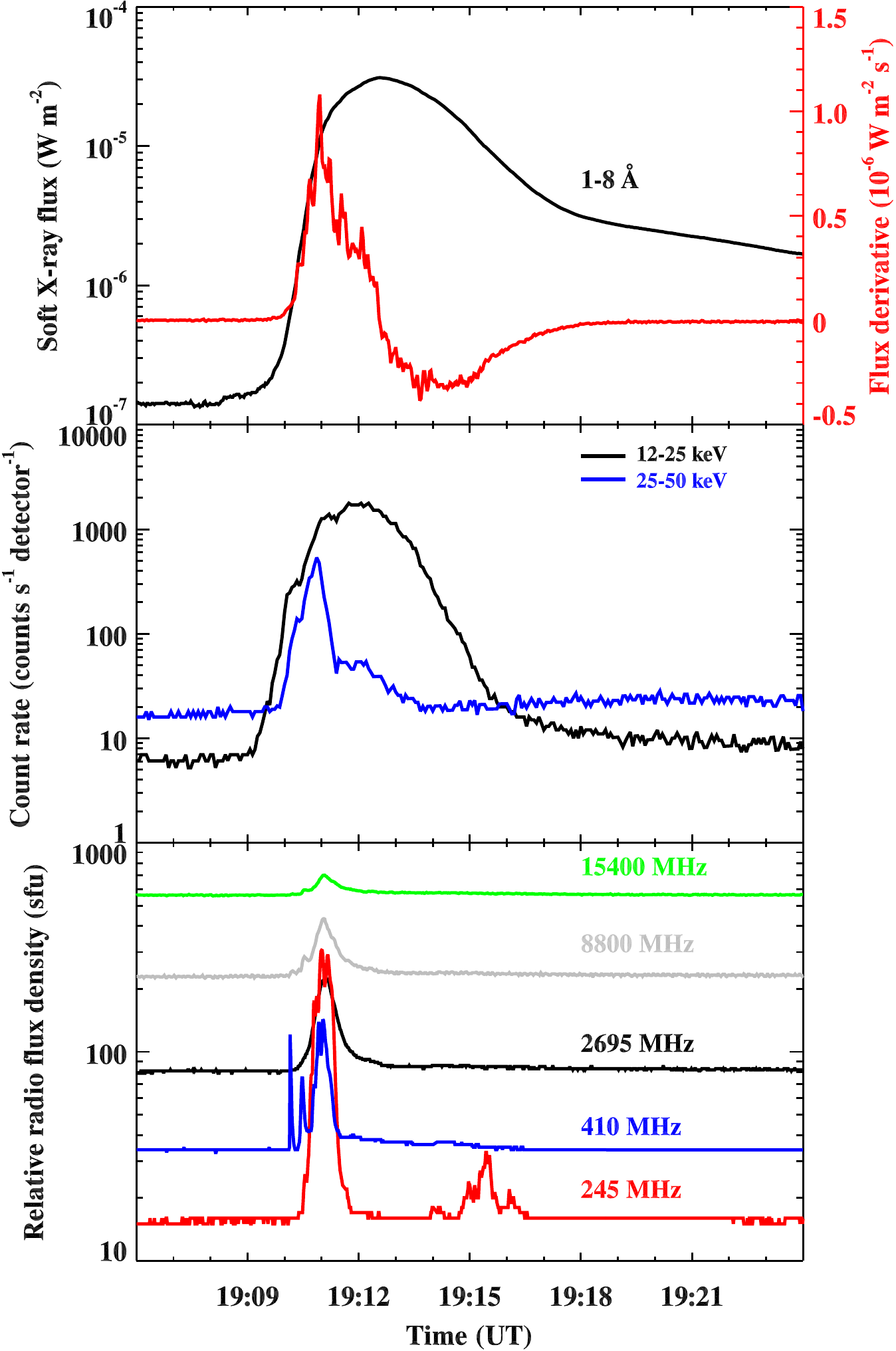}
}
\caption{GOES soft X-ray flux profile and its derivative (red), RHESSI hard X-ray flux profiles in 12-25 keV (black) and 25-50 keV (blue) energy bands, and radio fluxes in 245, 410, 2695, 8800, 15000 MHz from Sagamore-Hill station
on 16 October 2010.}
\label{flux}
\end{figure}
\begin{figure}
\centerline{
\includegraphics[width=0.55\textwidth]{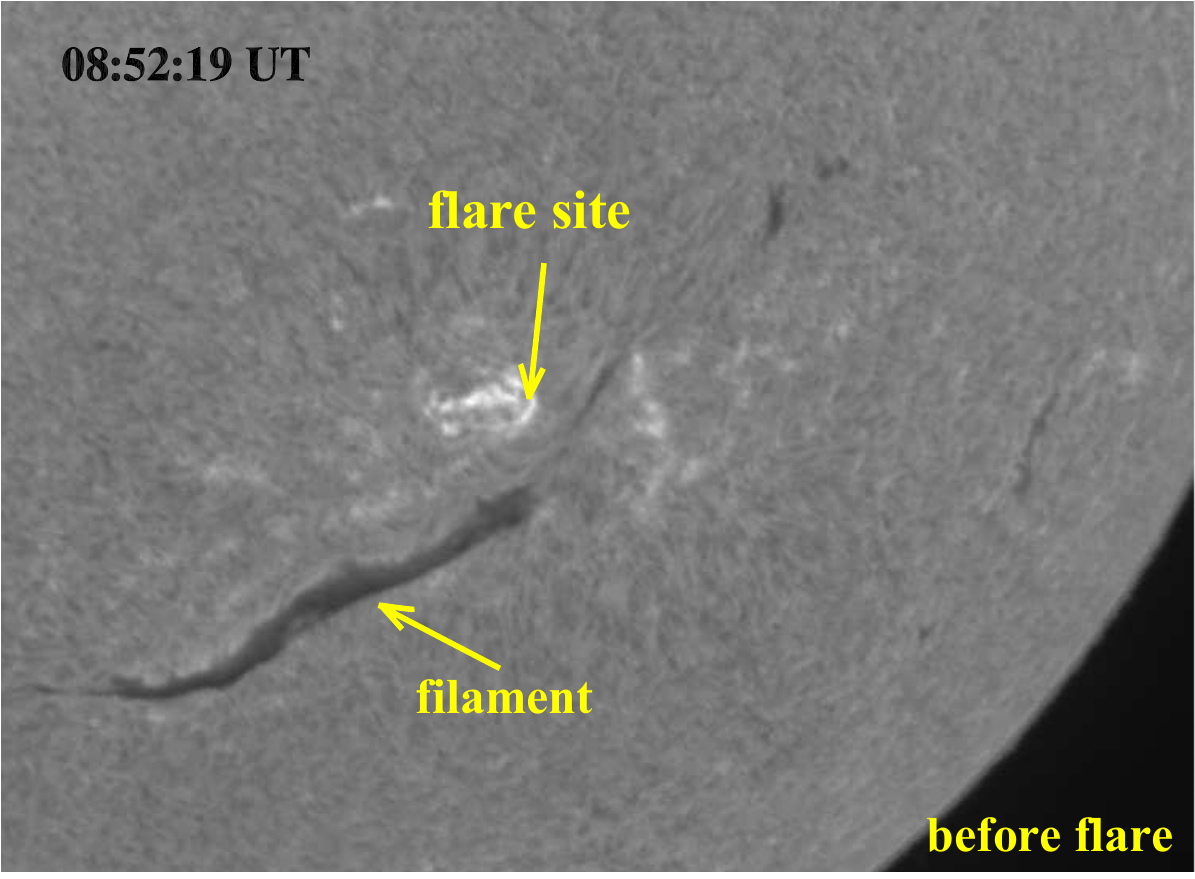}
\includegraphics[width=0.52\textwidth]{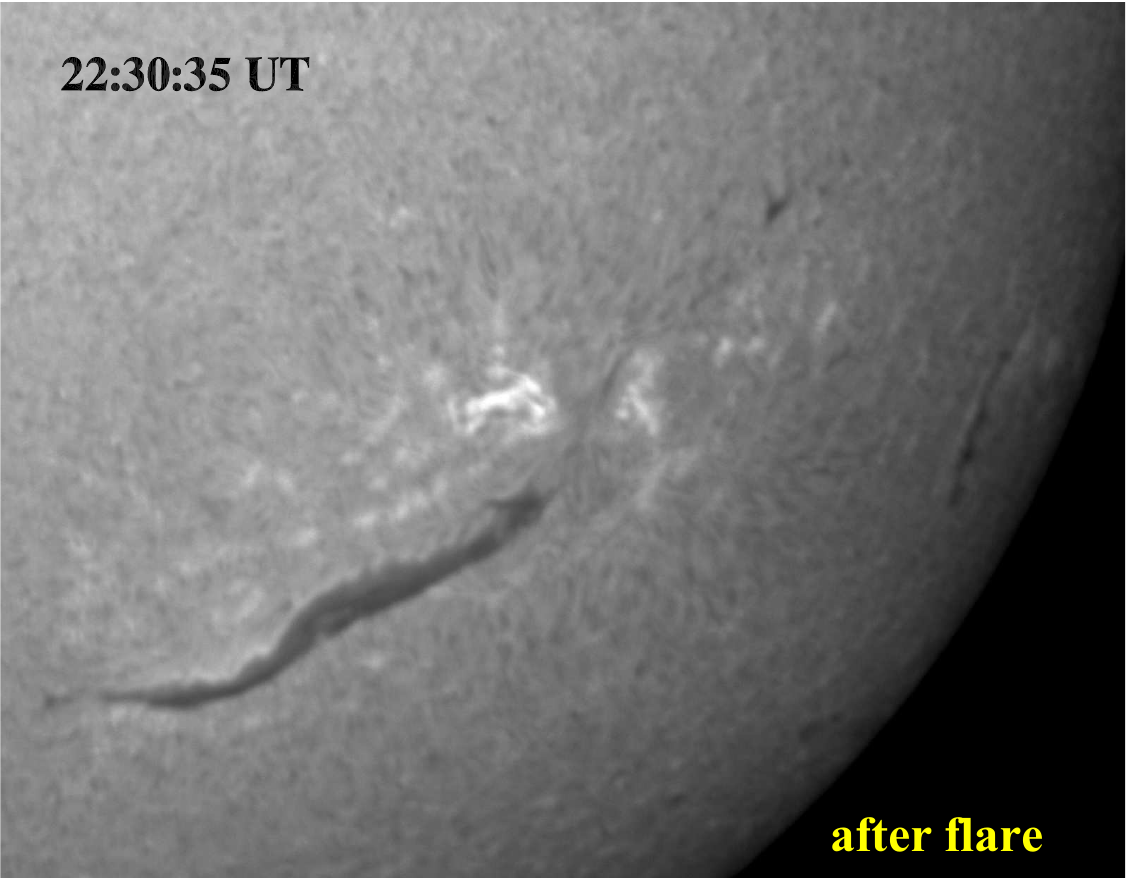}
}
\caption{H$\alpha$ images observed from BBSO and SMART telescopes respectively before (left) and after (right) the flare. It may be noted that there is no filament eruption during the flare.}
\label{halpha}
\end{figure}


\section{Multiwavelength Observations of M2.9/1N flare}
The active region NOAA 11112 ($\beta$-type configuration) was located in southern hemisphere (S19W29) on 16 October 2010. The first flare (M2.9/1N) associated with a slow CME was observed in this active region (AR) on 16 October 2010. Later, this AR produced  seven C-class flares during its passage through solar disk from 17-20 October 2010. Figure \ref{flux} displays the GOES soft X-ray flux (1-8 \AA) and its derivative, {\it Ramaty High Energy Solar Spectroscopic} Imager (RHESSI: \opencite{lin2002}) hard X-ray fluxes in 12-25 keV (black) and 25-50 keV (red), and radio flux profiles in different frequency bands observed at the Sagamore Hill station.
 An M2.9-class flare started at 19:07 UT, peaked at 19:12 UT and ended at 19:15 UT. It was a short-duration flare. In H$\alpha$, it was classified as 1N class flare, which started at 19:10 UT, maximized at 19:13 UT and ended at 19:30 UT. The soft X-ray flux derivative matches the hard X-ray flux specially in 25-50 keV, which suggests that the same electrons are responsible for both thermal (SXR) and nonthermal (HXR) emissions, {\it i.e.}, the Neupert effect holds \cite{neupert1968,veronig2002}. 
The radio flux profiles in different frequencies show a peak at 19:11 UT, which correlates well with the hard X-ray emission. The flux profile in 410 MHz shows two peaks prior to the main peak (19:11 UT). This may be related to the precursor of the flare and may correspond to the typical height of the reconnection. The decimetric bursts (100-4000 MHz) match with the hard X-ray emission, which are believed to be associated with the  nonthermal electrons emitted by a coherent process \cite{benz2005}. Coherent radio
emission at decimeter wavelengths originates in either a plasma
density between 3$\times$10$^{8}$ and 3$\times$10$^{10}$ cm$^{3}$ or a magnetic field
between 100 and 1000 G (assuming second harmonic emission
is present), which are typically expected in flare
acceleration regions \cite{dabr2009}.

Figure \ref{halpha} displays the H$\alpha$ images of the active region before (08:52:19 UT) and after (22:30:35 UT) the flare observed respectively at Big Bear Solar Observatory (BBSO) and with the {\it Solar Magnetic Activity Research Telescope} (SMART) \cite{ueno2004}. The active region was associated with a huge filament lying along southeast direction. The flare occurred not far from the filament, which did not erupt during the flare event.
\begin{figure}
\centerline{
\includegraphics[width=0.5\textwidth]{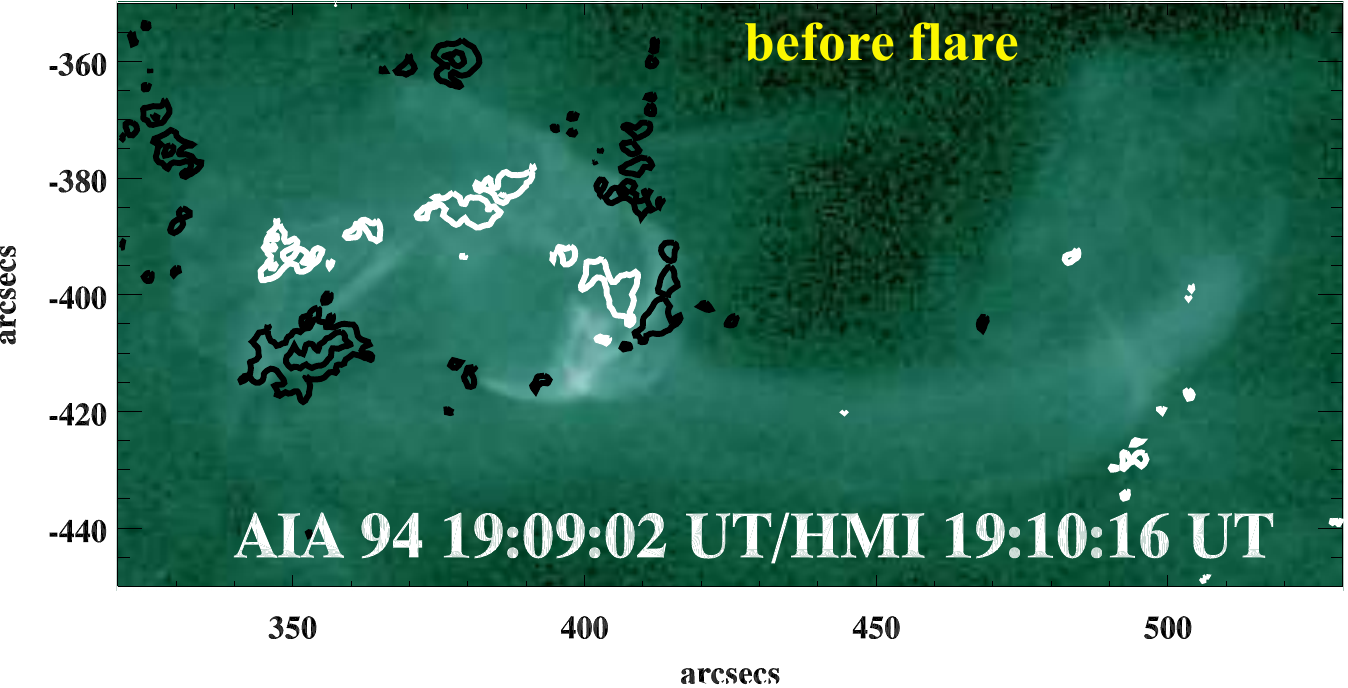}
\includegraphics[width=0.5\textwidth]{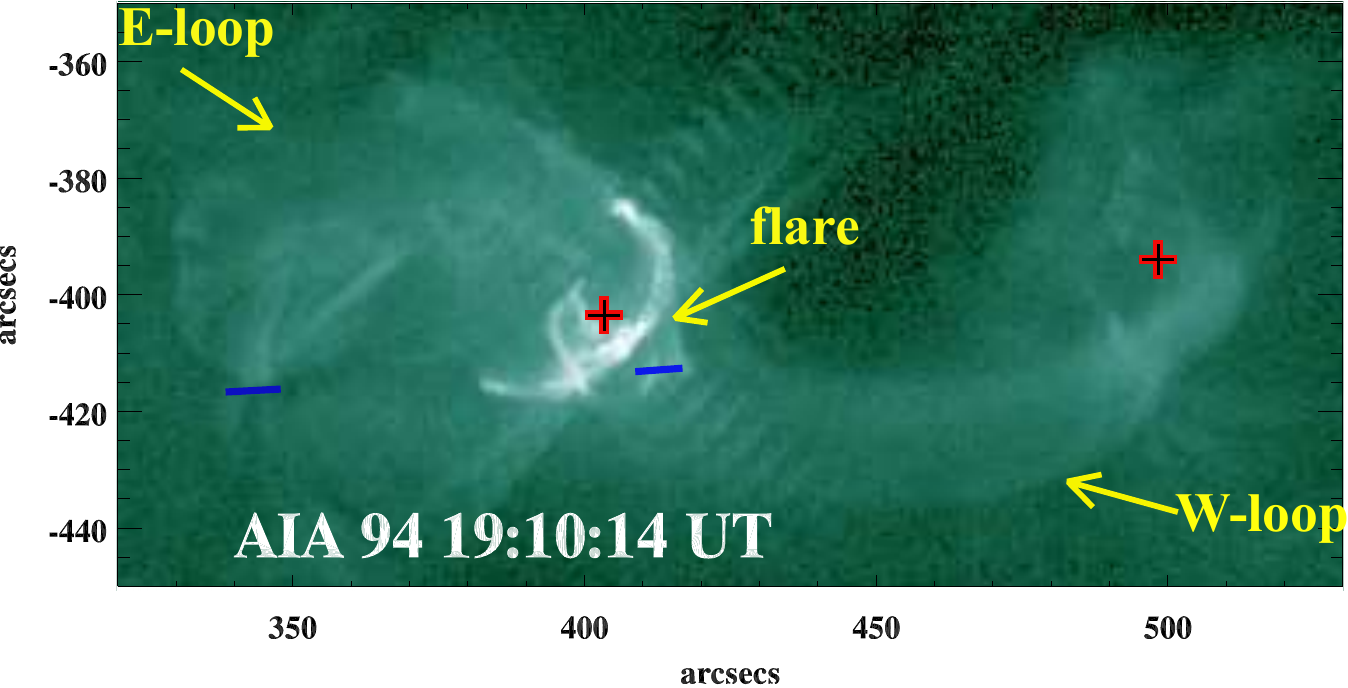}
}
\centerline{
\includegraphics[width=0.5\textwidth]{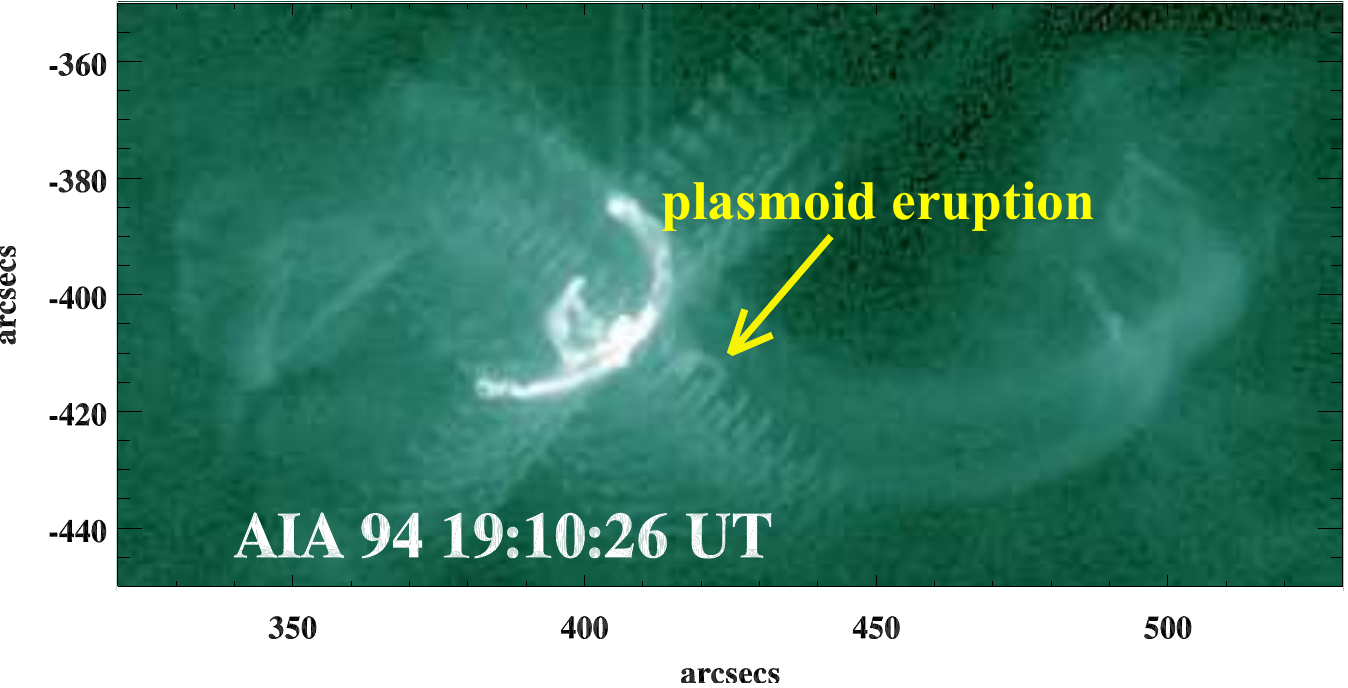}
\includegraphics[width=0.5\textwidth]{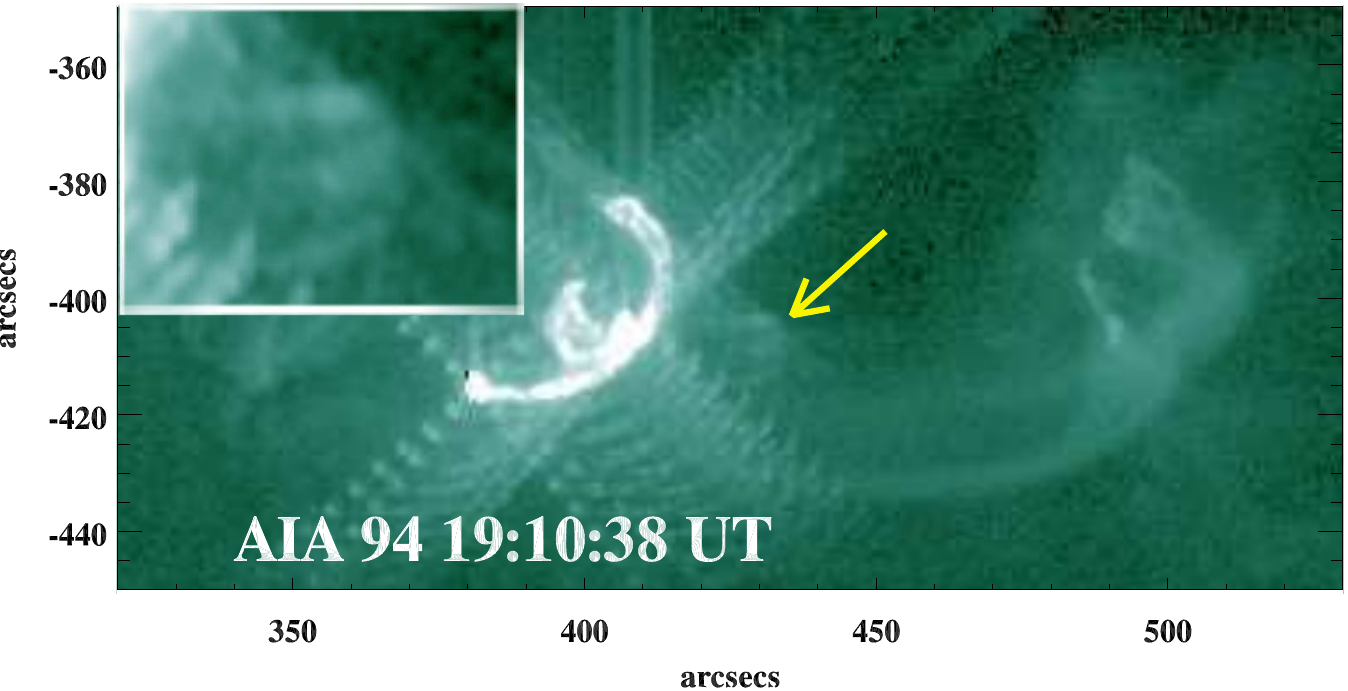}
}
\centerline{
\includegraphics[width=0.5\textwidth]{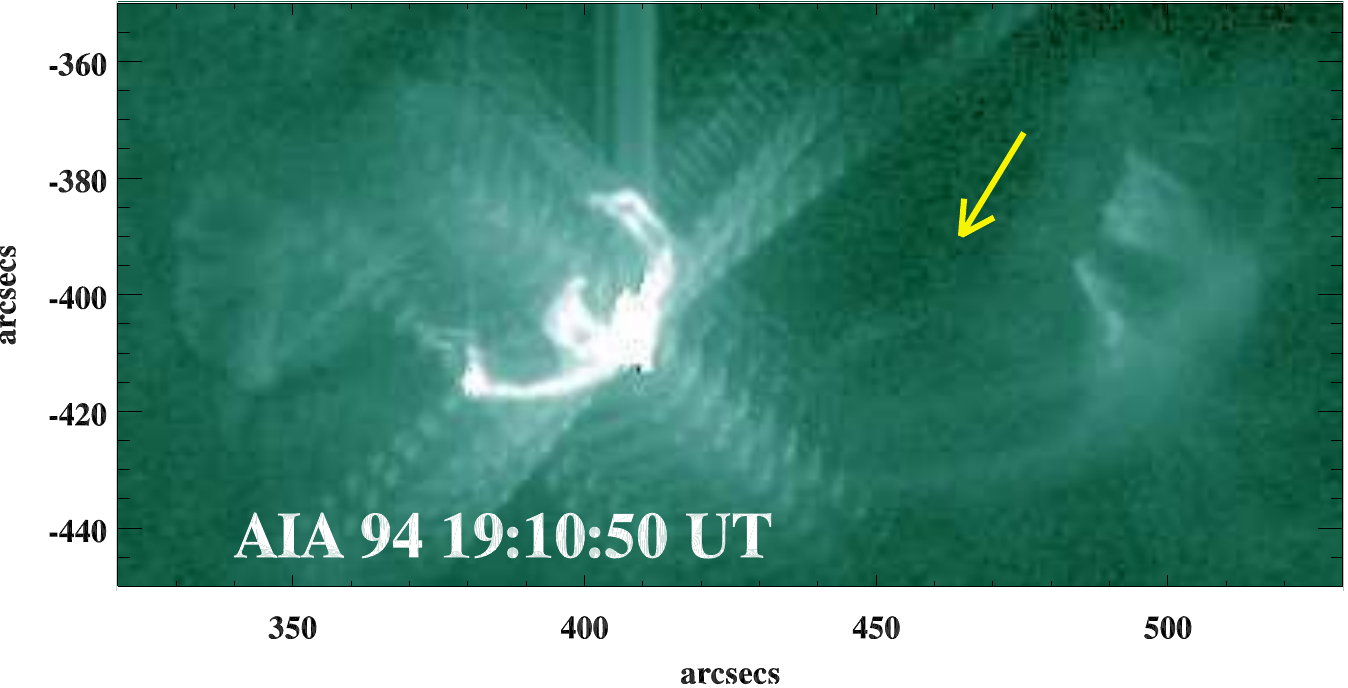}
\includegraphics[width=0.5\textwidth]{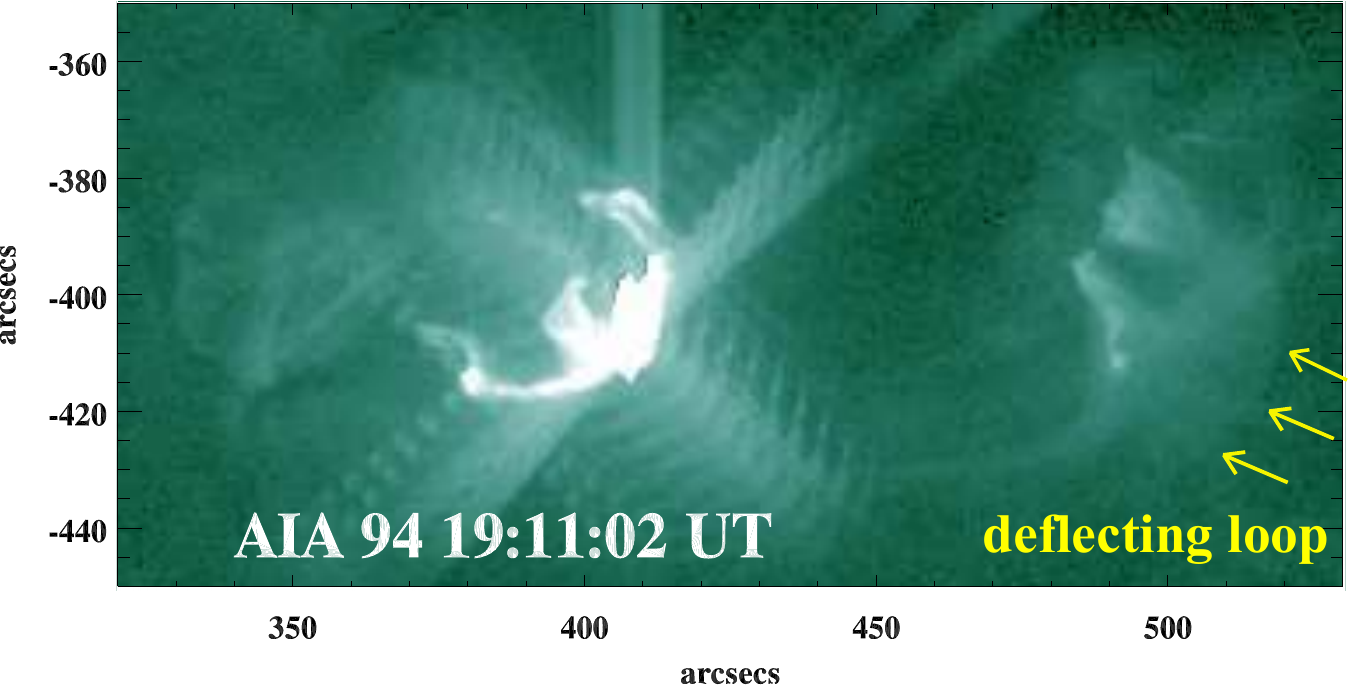}
}
\centerline{
\includegraphics[width=0.5\textwidth]{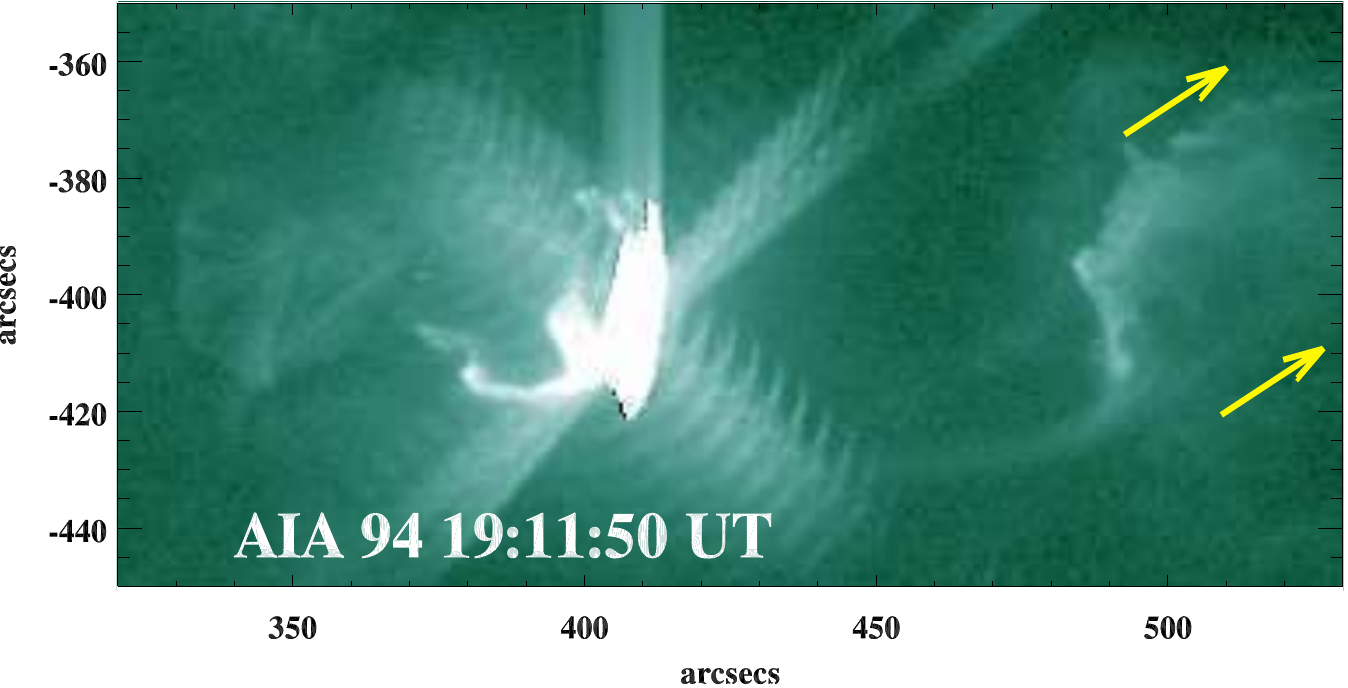}
\includegraphics[width=0.5\textwidth]{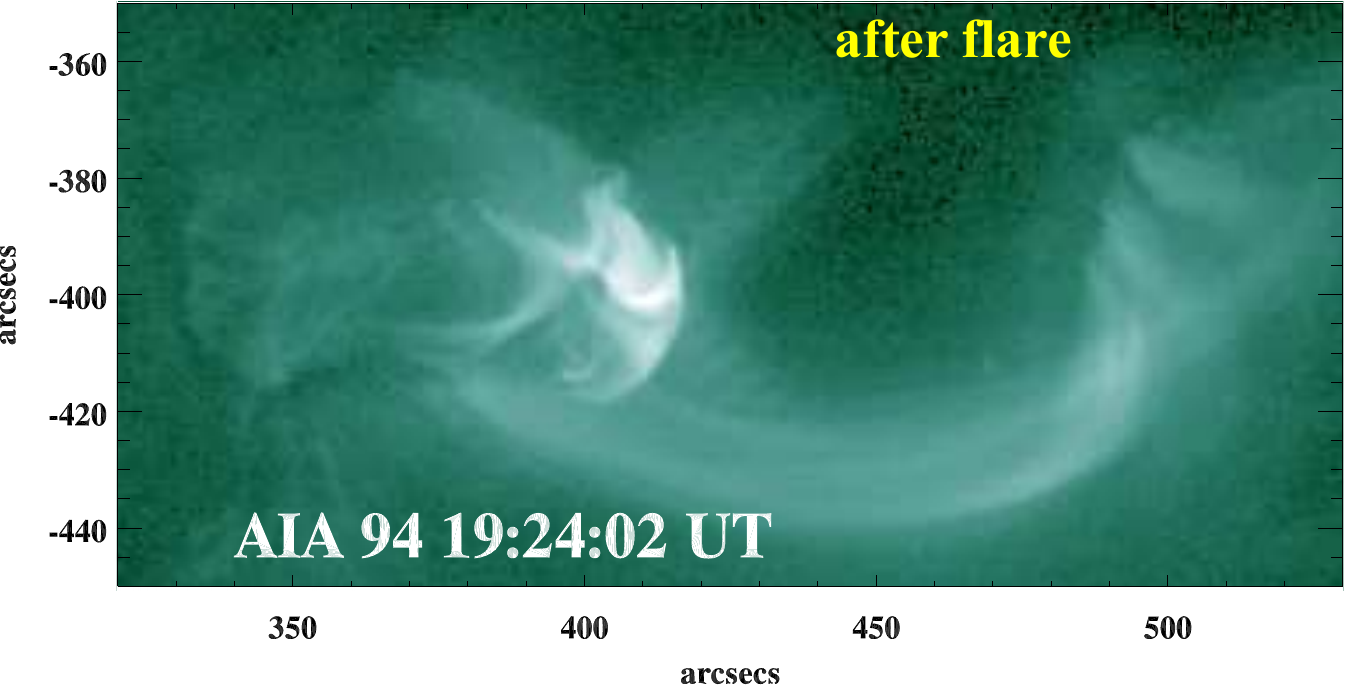}
}
\caption{Selected SDO/AIA 94 \AA \ EUV images showing the initiation of M2.9 flare associated with plasmoid eruption and deflection of loop system on 16 October 2010. The top-left image has been overlaid by HMI magnetogram contours of positive (white) and negative (black) polarities. The contour levels are $\pm$500, $\pm$1000, $\pm$2000 and $\pm$3000 gauss (G). The enlarged view of the plasmoid is shown in the top-left corner of a panel at 19:10:38 UT.}
\label{aia}
\end{figure}
\begin{figure}
\centerline{
\includegraphics[width=0.7\textwidth]{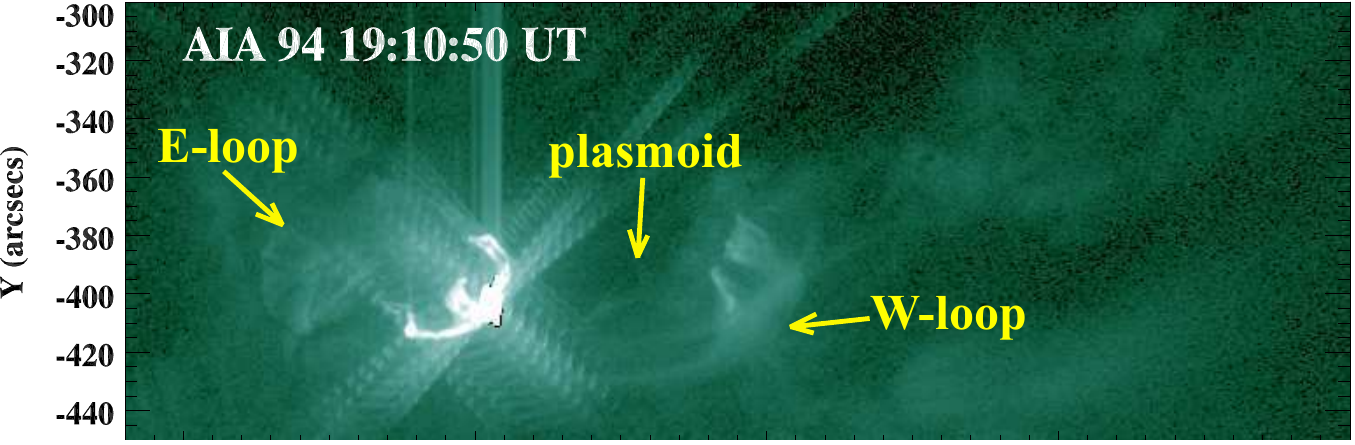}
}
\centerline{
\includegraphics[width=0.7\textwidth]{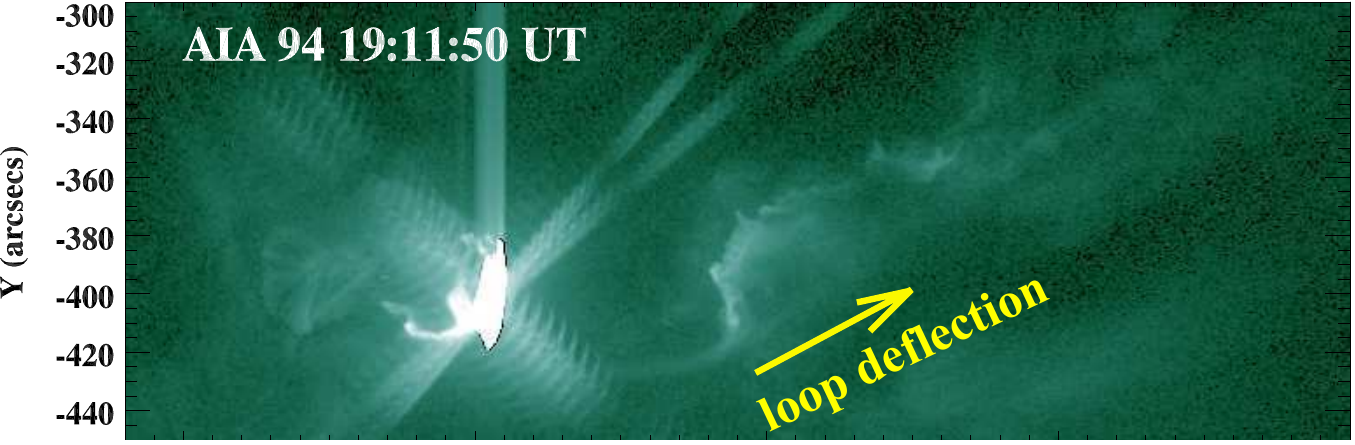}
}
\centerline{
\includegraphics[width=0.7\textwidth]{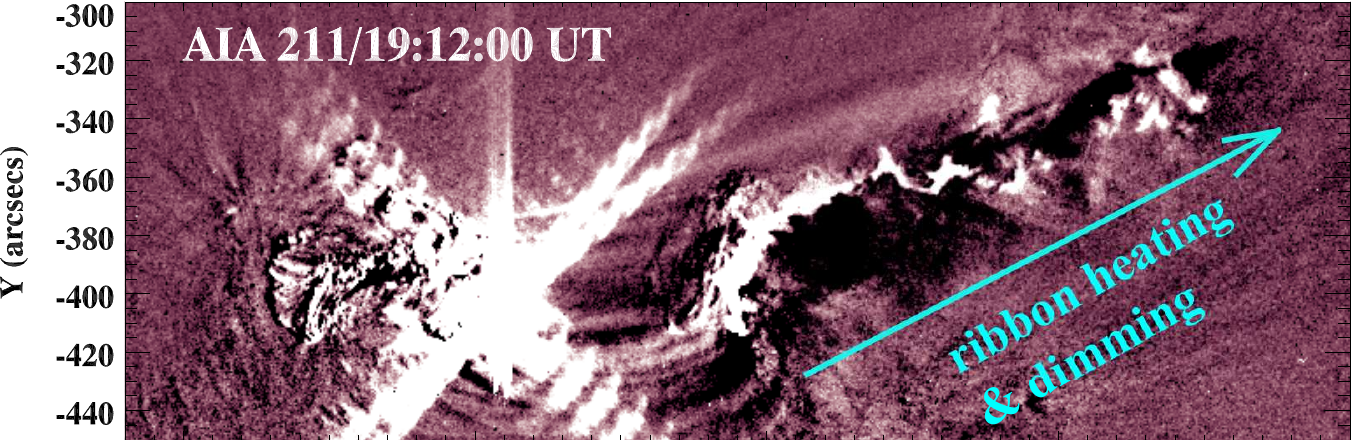}
}
\centerline{
\includegraphics[width=0.7\textwidth]{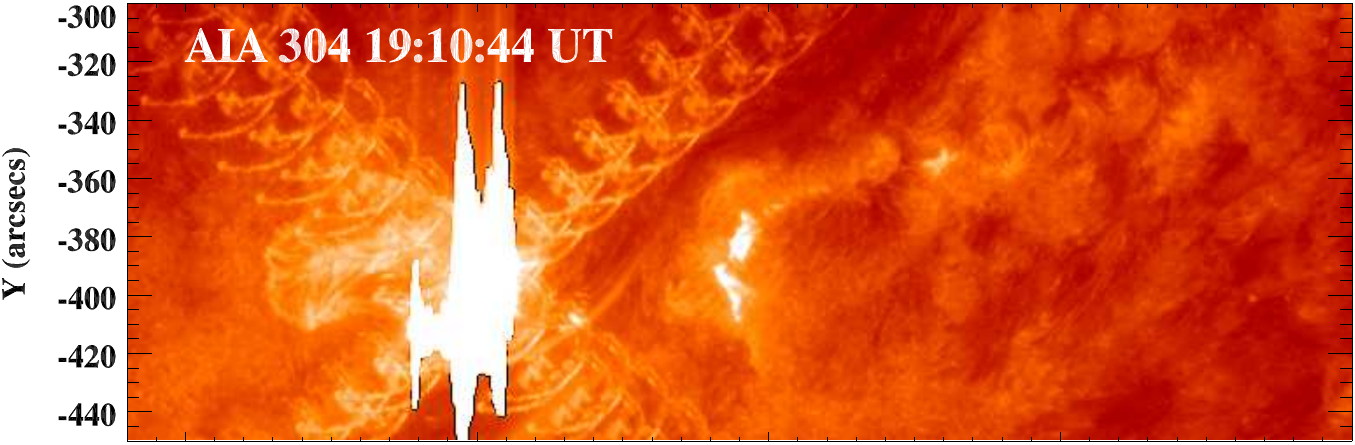}
}
\centerline{
\hspace{0.06cm}
\includegraphics[width=0.715\textwidth]{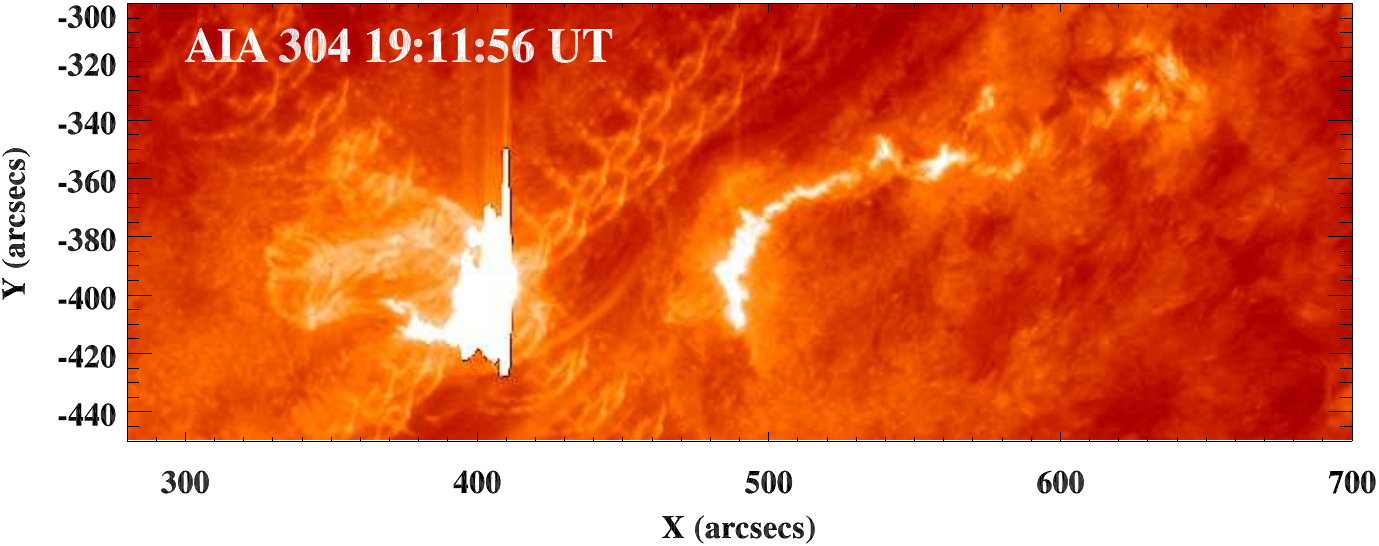}
}
\caption{Selected SDO/AIA 94, 211 (base difference) and 304 \AA \ EUV images showing the deflection of the W-loop by a westward propagating disturbance and associated extended ribbon brightening.}
\label{ribbon}
\end{figure}


 SDO/{\it Atmospheric Imaging Assembly} (AIA: \opencite{lemen2012}) observes multiwavelength images at 0.6$^{\prime\prime}$ per pixel resolution with 12 s cadence. The field of view is 1.3$R_\odot$, so it covers the full solar disk. We use the AIA 211 \AA \ (Fe {\small XIV}, $T\approx$2 MK), 304 \AA \ (He {\small II}, $T\approx$0.05 MK), 1700 \AA \ (T=5000 K) and 94 \AA \ (Fe {\small XVIII}, $T\approx$6.3 MK) images, which cover the height from the photosphere to the corona. 
 Figure \ref{aia} displays SDO/AIA 94 \AA \ EUV images, revealing the coronal configuration of the active region and associated flare brightening. These images are favorable to observe the evolution of high temperature plasma in the flaring region. The top-left panel has been overlaid by {\it Helioseismic and Magnetic Imager} (HMI: \opencite{schou2012}) magnetogram contours of positive (white) as well as negative polarities (black). This shows the connectivity of the footpoints of the corresponding loop-systems before the flare. Careful inspection of these images reveals the existence of two major loop systems in the active region. The top-right image shows one longer loop system in the westward direction (indicated by ``W-loop"), whereas a smaller loop system was protruding in the eastward direction (indicated by ``E-loop"). The footpoint polarities of these loops are indicated by `+' and `-' symbols. The flare brightening started at the western footpoint of the E-loop system in between the opposite-polarity regions (indicated by arrow). 
 
A plasmoid ejection was also observed during the flare impulsive phase (19:10:26 UT to 19:10:50 UT). The panel at 19:10:26 UT shows the plasmoid ejection indicated by arrow. The moving plasmoid structure is revealed in the panels at 19:10:38 UT and 19:10:50 UT (marked by arrows). We measured the distance of the plasmoid's leading edge from the flare center and calculated the speed from the linear fit to the projected height-time measurements, which was $\approx$1197 km s$^{-1}$. Moreover, we observe a strong deflection at the right footpoint of the W-loop system just after the plasmoid's disappearance during the flare peak time at 19:11:02 UT (please see AIA 94 \AA \ movie). The deflection at the footpoint of the loop system is likely caused by the propagation large scale coronal wave away from the flare site, which is studied in detail in \inlinecite{kumar2012b}(hereafter Paper II). After the flare (19:24:02 UT), we can see the two loop systems as observed prior to the flare event. It should be noted that these two loop-systems are filled with hot plasma as evident in 94 \AA \ images. Therefore, AIA images show the coronal environment at the flare site, associated flare brightening, plasmoid ejection as well as rearrangement of the loop systems.
Most of the nonthermal emission in hard X-ray (25-50 keV) as well as in radio flux profile peaked at 19:11 UT (Figure \ref{flux}), during the acceleration phase of the plasmoid. This scenario conforms with the standard eruptive flare model (i.e. CSHKP), where the nonthermal emission matches well with the timing of plasmoid acceleration \cite{ohyama1998,yok2001,maricic2007,temmer2008,temmer2010}.  

\begin{figure}
\centerline{
\includegraphics[width=0.5\textwidth]{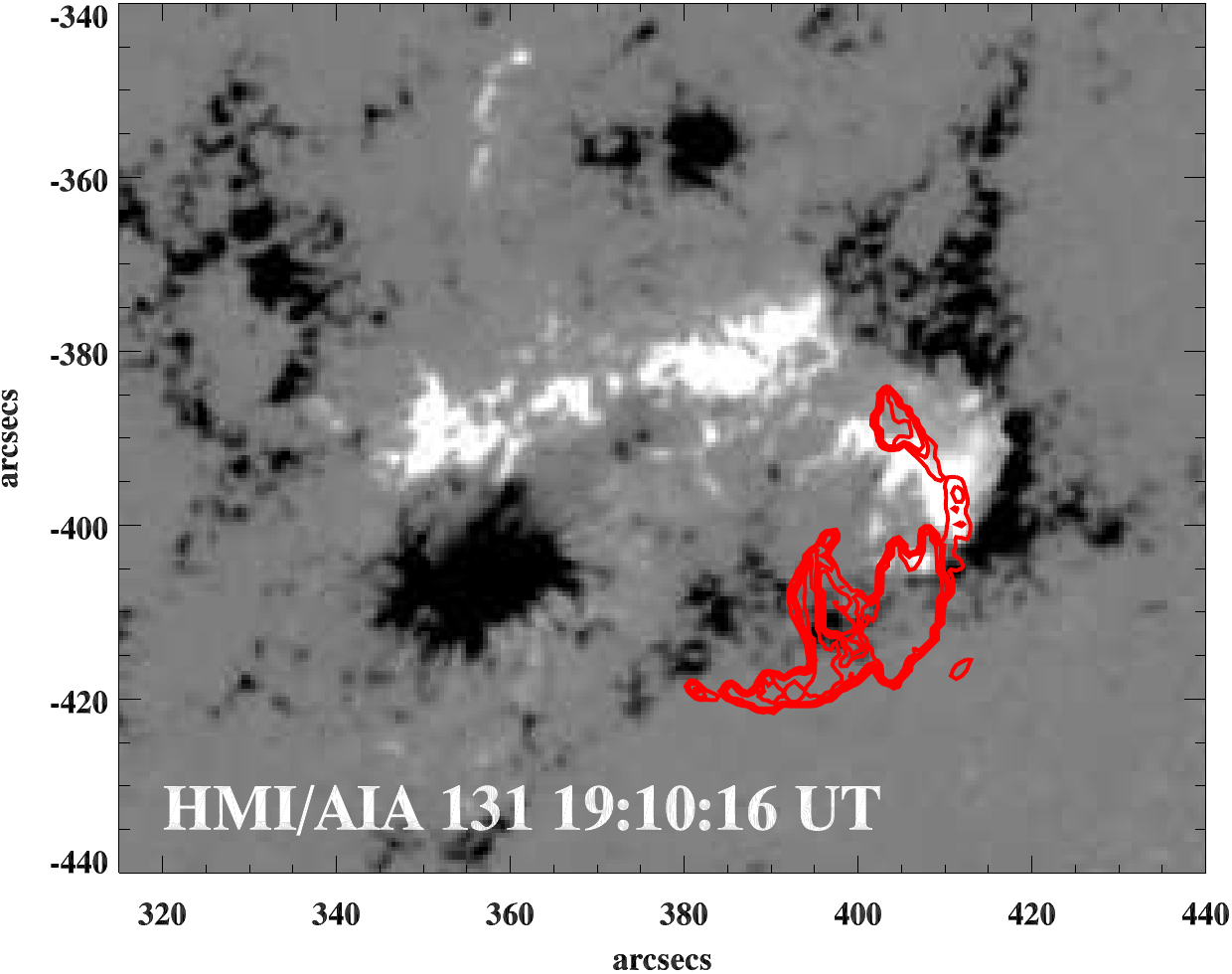}
\thicklines
$\color{blue} \put(-45,62){\circle{140}}$
\includegraphics[width=0.5\textwidth]{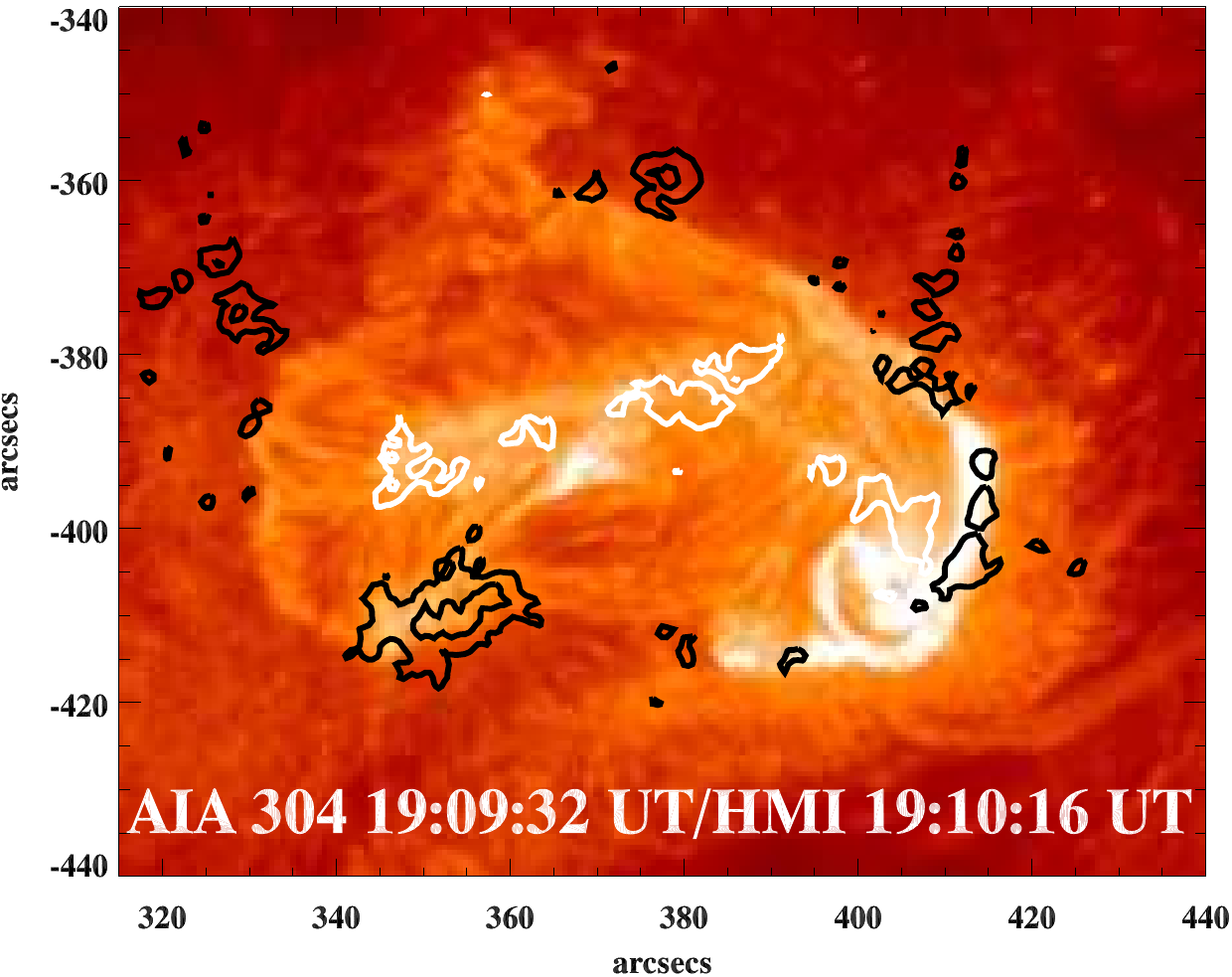}
$\color{blue} \put(-50,62){\circle{140}}$
}
\caption{Left: SDO/HMI magnetogram overlaid by AIA 131 \AA \ image intensity contours. The contour levels are 30$\%$, 50$\%$, 70$\%$, 90$\%$ of the peak intensity. Right: HMI magnetogram contours overlaid on AIA 304 \AA \ image during the initial phase of the flare.} The contour levels are $\pm$500, $\pm$1000 G. White/Black contours indicate positive/negative polarities. 
\label{aia_hessi}
\end{figure}

 In order to illustrate the deflection of the W-loop and associated coronal and chromospheric responses of the flare, we have shown some of the selected AIA images in 94, 211 and 304 \AA \ in Figure \ref{ribbon}.  AIA 94 \AA \ images show the westward deflection of the W-loop (indicated by an arrow), just after the plasmoid ejection (top two panels). We can see the extended ($\approx$200$^{\prime\prime}$) ribbon brightening in AIA 211 \AA \ base difference image at 19:12 UT. The flare ribbon is also observed in AIA 304 \AA \ (i.e. chromosphere and transition region) images. In AIA 94 and 211 \AA \ movies, we see the propagating disturbance in the western direction starting from the flare center. The coronal dimming along the direction of propagating brightening probably indicates the depletion of density. The deflection of the coronal W-loop and ribbon brightening may be associated with a coronal wave propagating from the flare site. A detailed study of the coronal wave is presented in Paper II.

Figure \ref{aia_hessi} displays HMI magnetogram and AIA 304 \AA~ image.
The left image shows the HMI magnetogram overlaid by AIA 131 \AA~ flare intensity contours. The contour levels are 30$\%$, 50$\%$, 70$\%$, and 90$\%$ of the peak intensity. It is noted that the flare brightening in corona was located above a pair of small opposite polarity field regions (shown within the circle). The right image displays HMI magnetogram contours over AIA 304 \AA \ image during the flare initiation. AIA 304 \AA \ image corresponds to upper chromosphere as well as the transition region of the solar atmosphere. The initial brightening started in between the opposite polarity bipolar region as indicated within blue circle. 
\begin{figure}
\centerline{
\includegraphics[width=0.33\textwidth]{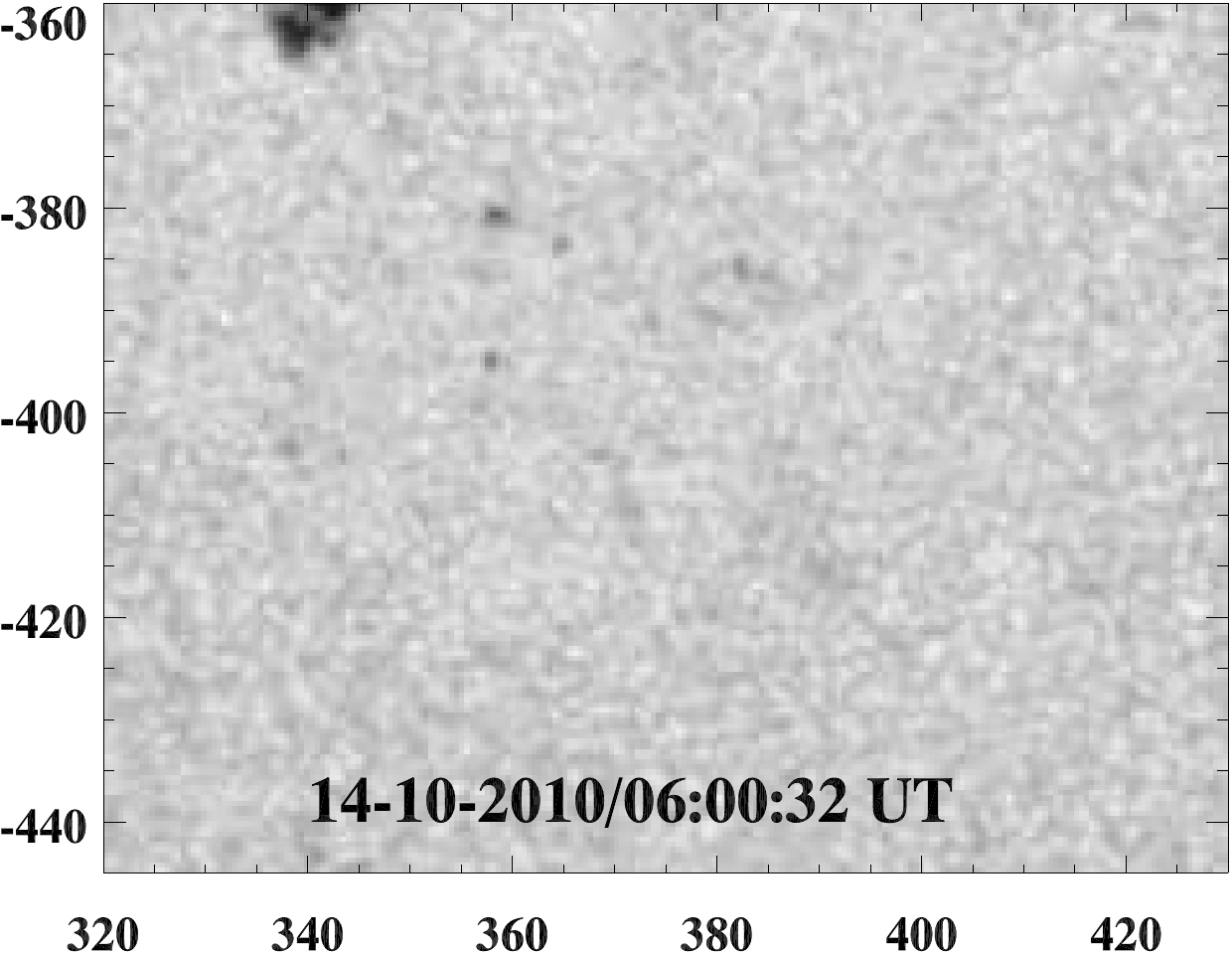}
\includegraphics[width=0.33\textwidth]{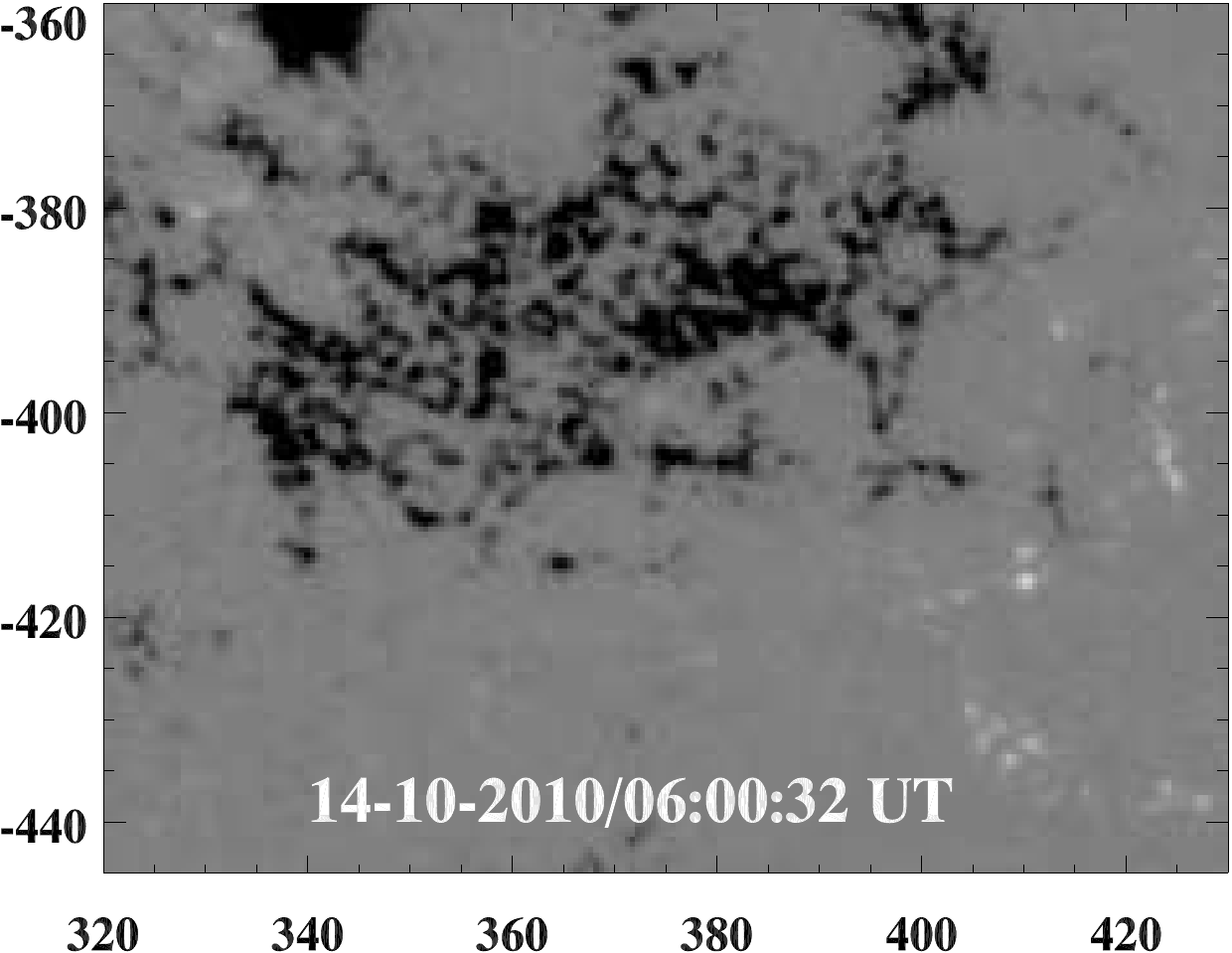}
\includegraphics[width=0.33\textwidth]{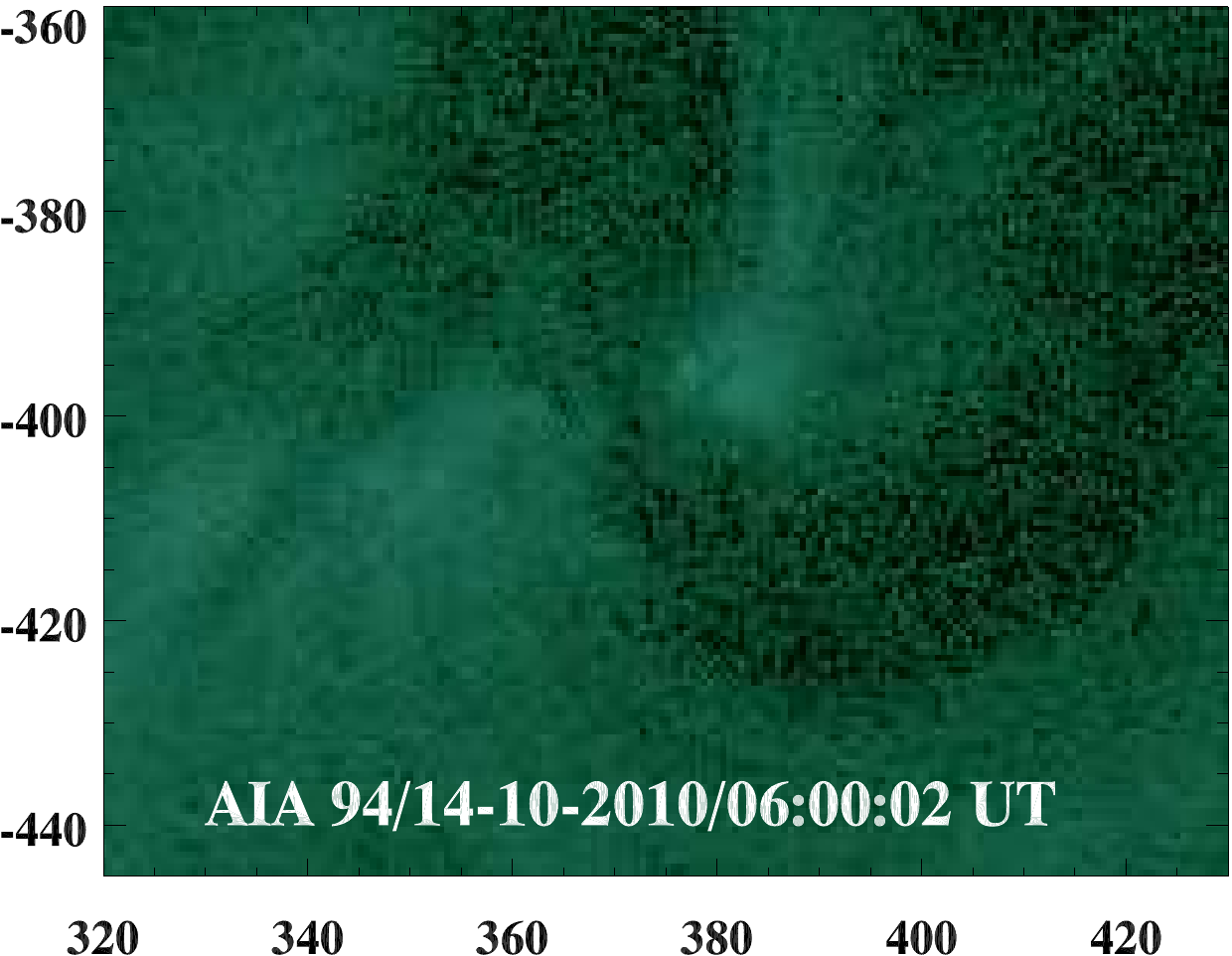}
}
\centerline{
\includegraphics[width=0.33\textwidth]{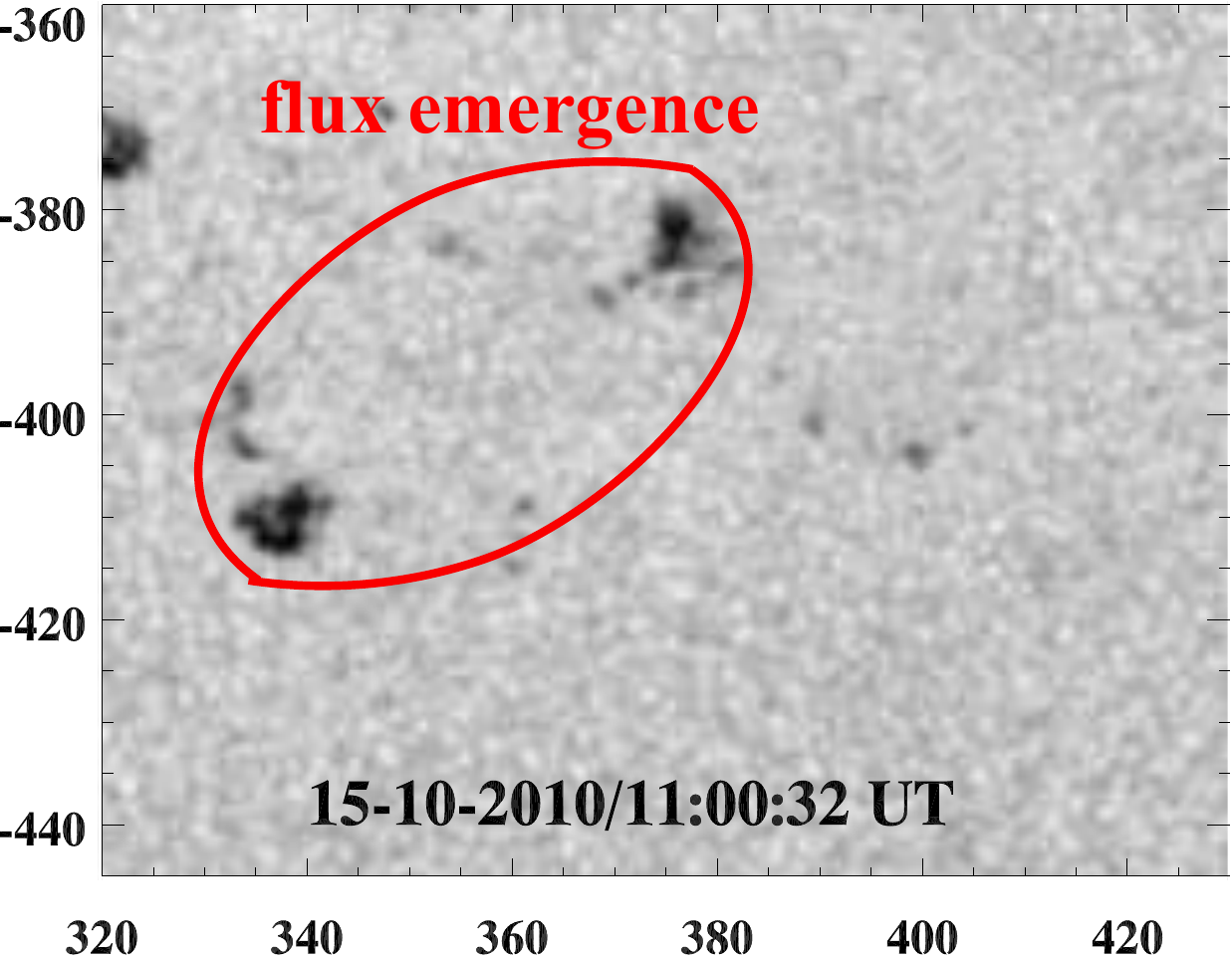}
\includegraphics[width=0.33\textwidth]{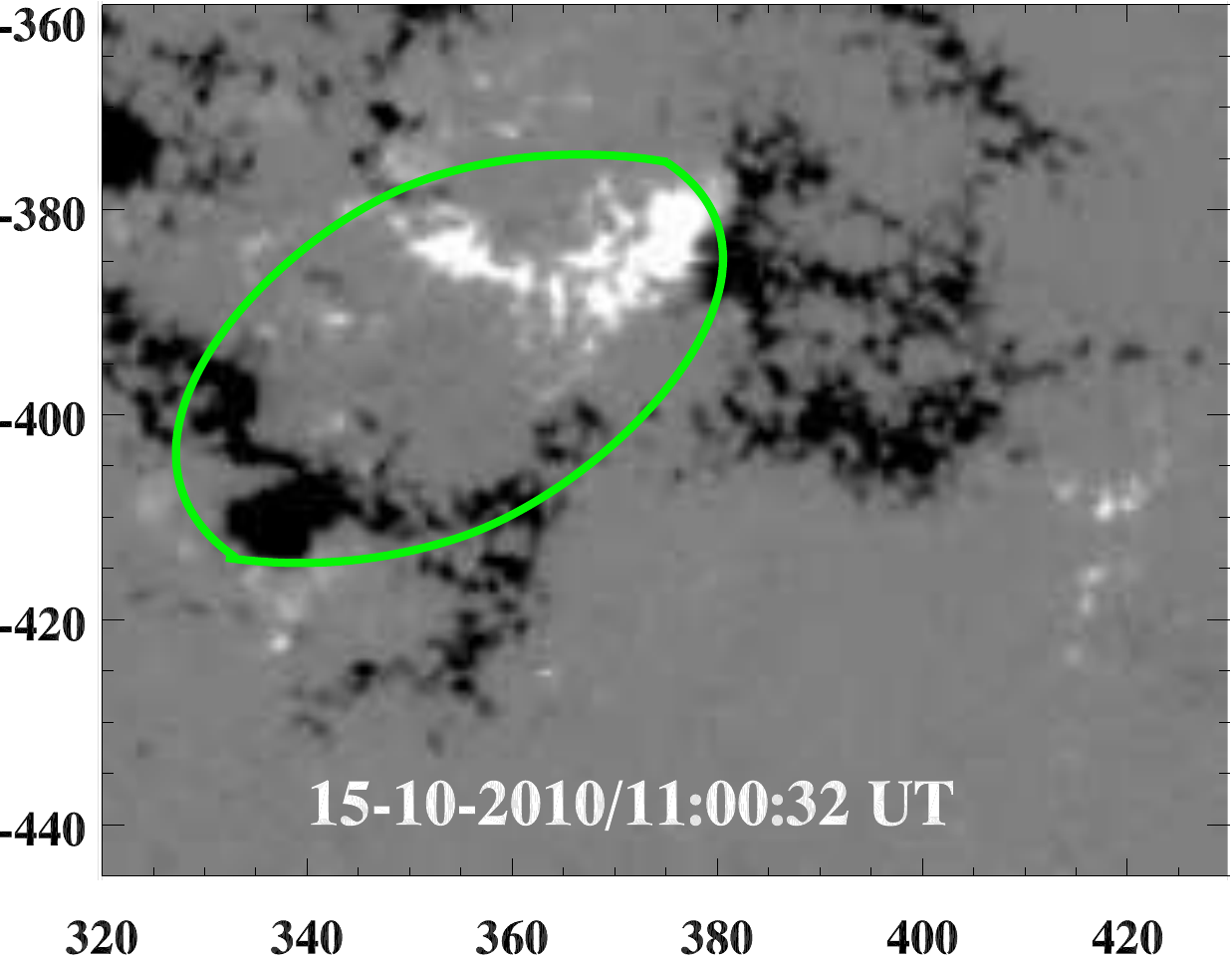}
\includegraphics[width=0.33\textwidth]{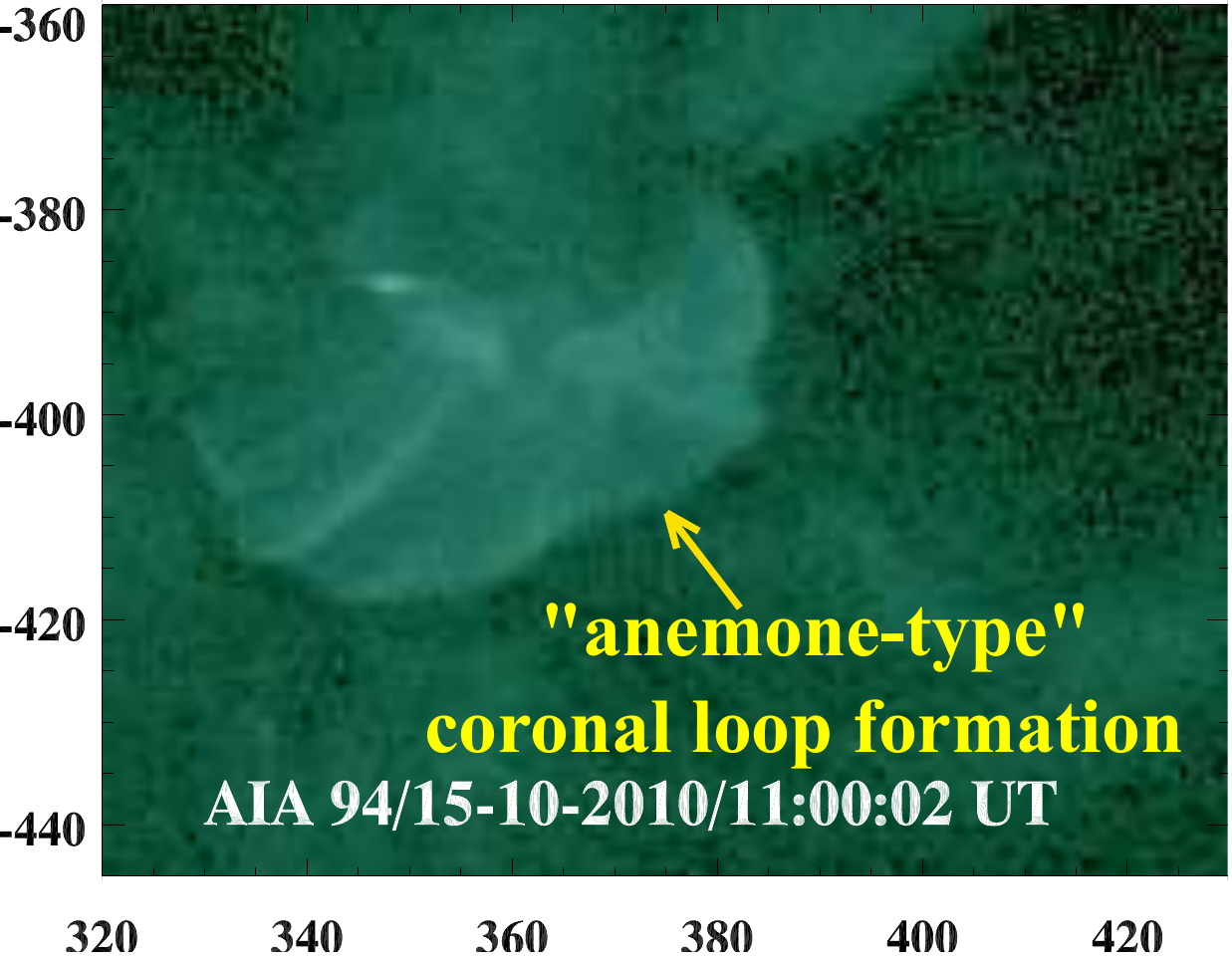}
}
\centerline{
\includegraphics[width=0.33\textwidth]{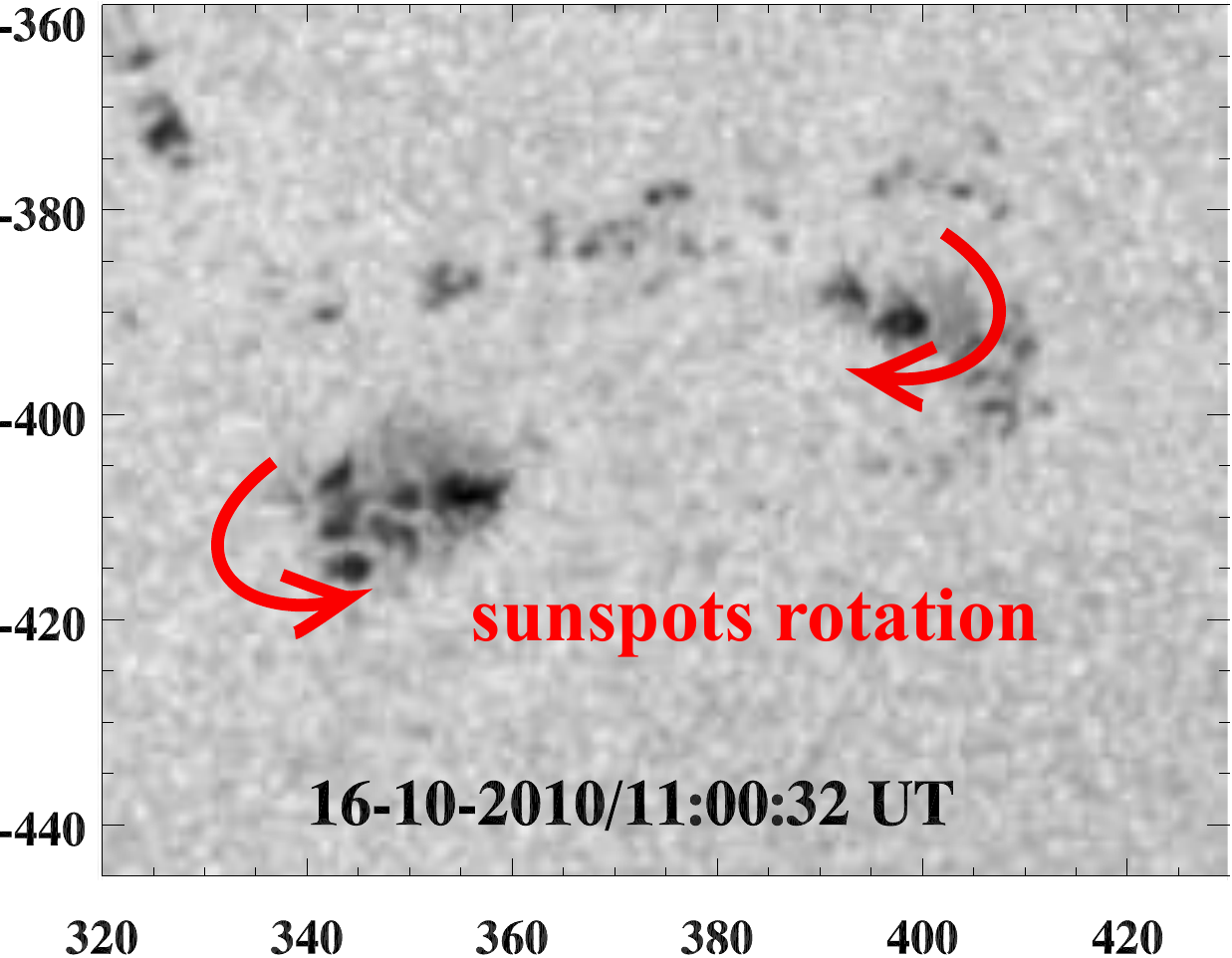}
\includegraphics[width=0.33\textwidth]{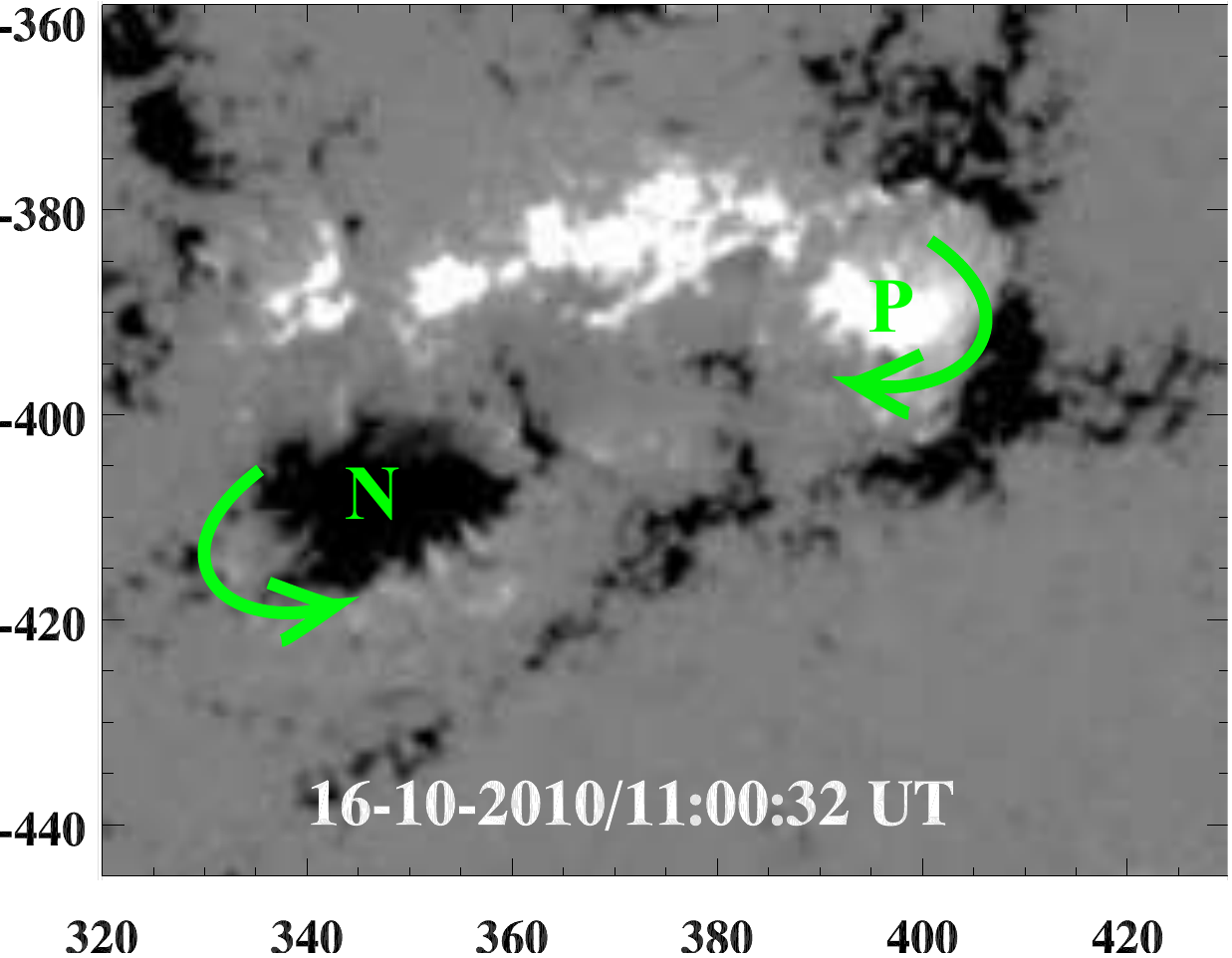}
\includegraphics[width=0.33\textwidth]{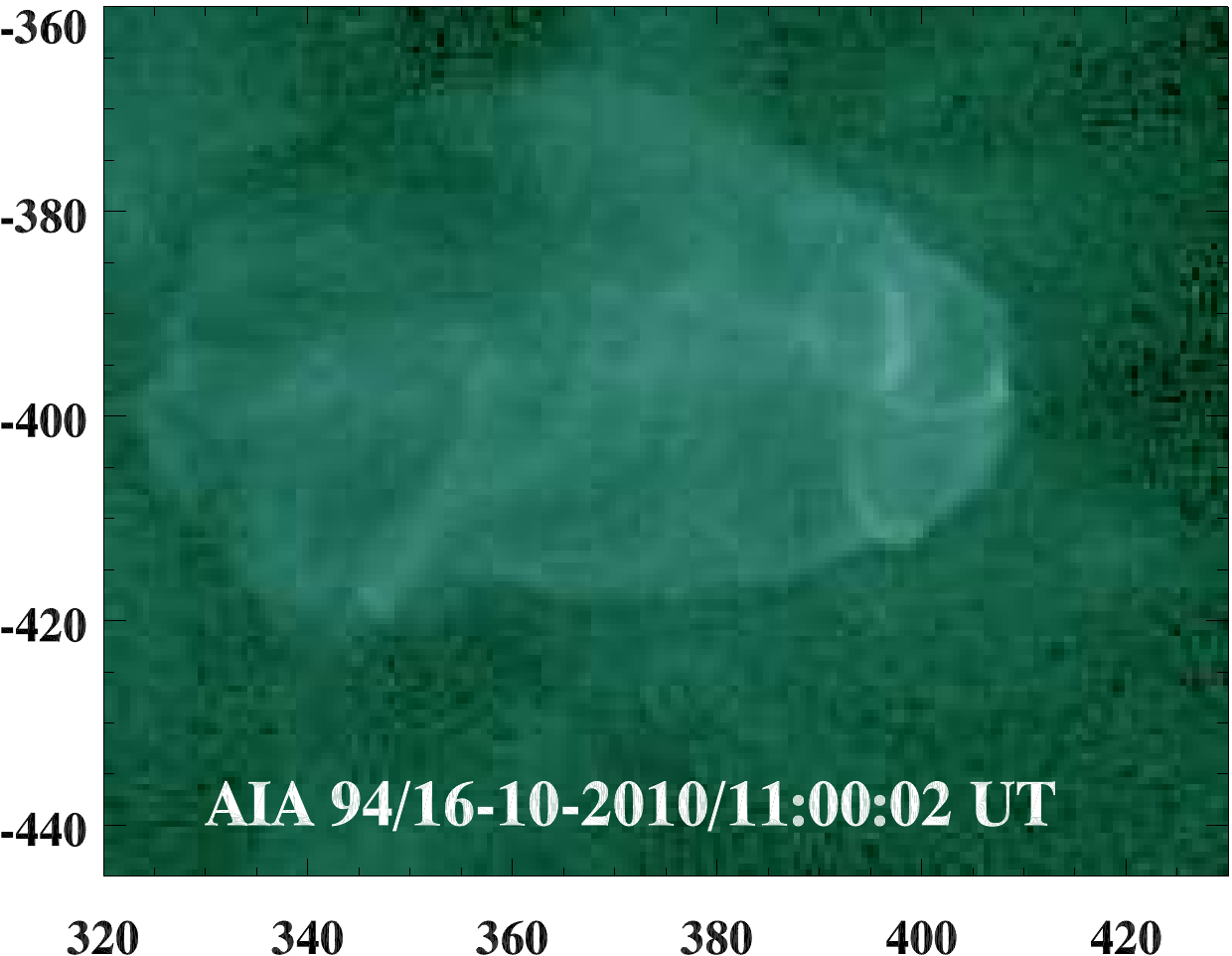}
}
\centerline{
\includegraphics[width=0.33\textwidth]{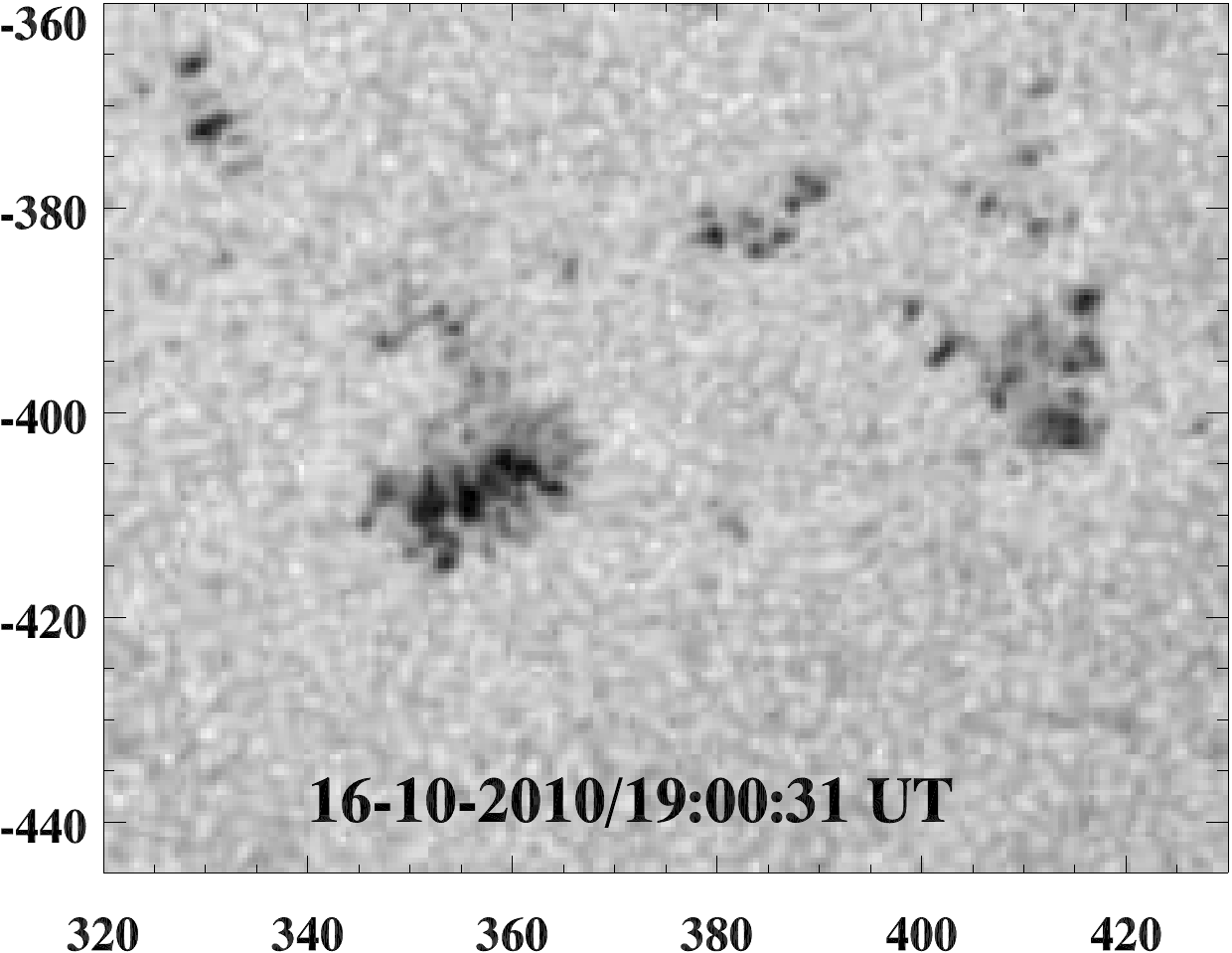}
\includegraphics[width=0.33\textwidth]{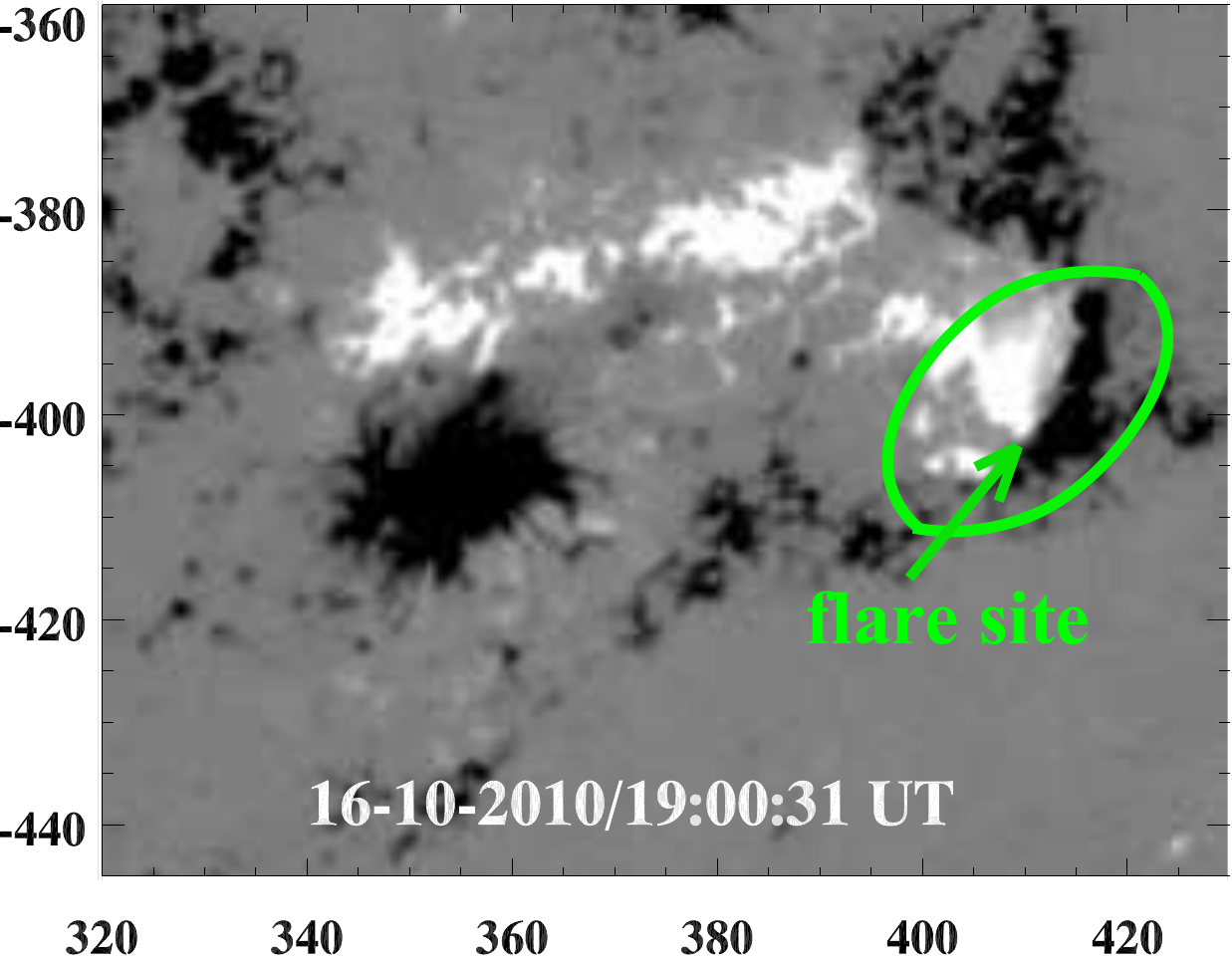}
\includegraphics[width=0.33\textwidth]{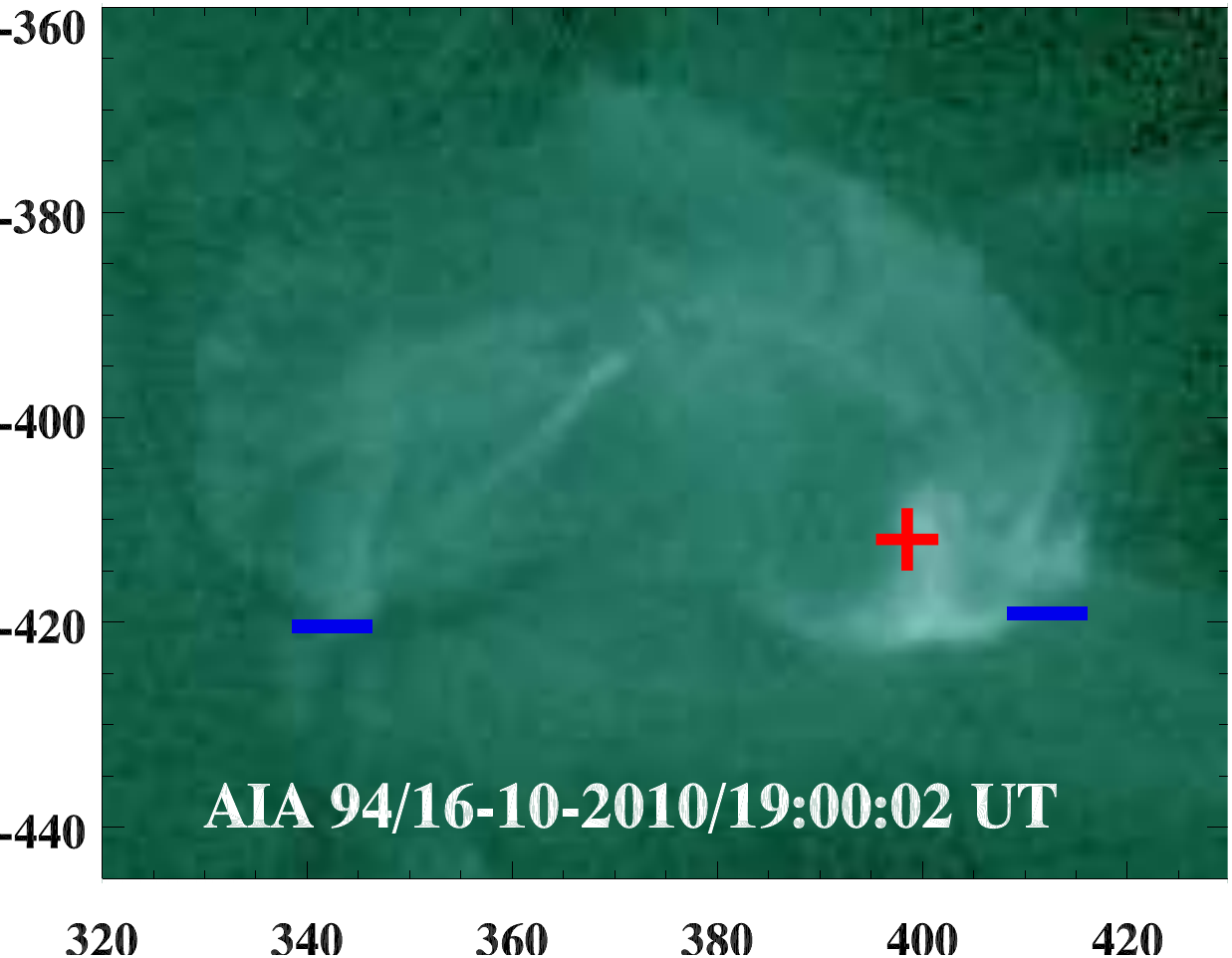}
}
\caption{HMI continuum, magnetograms and AIA 94 \AA \ EUV images before the flare occurrence. X and Y axis are in arcsecs.}
\label{cont}
\end{figure}
\begin{figure}
\centerline{
\includegraphics[width=0.5\textwidth]{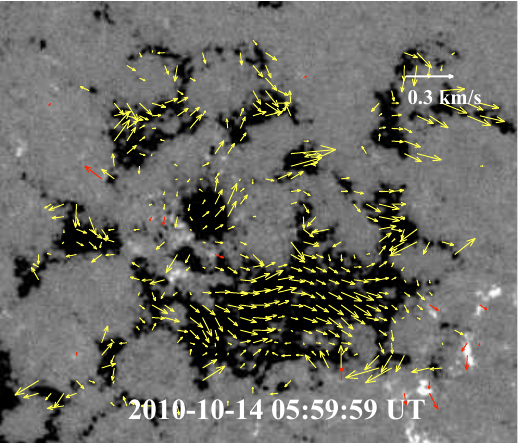}
\includegraphics[width=0.5\textwidth]{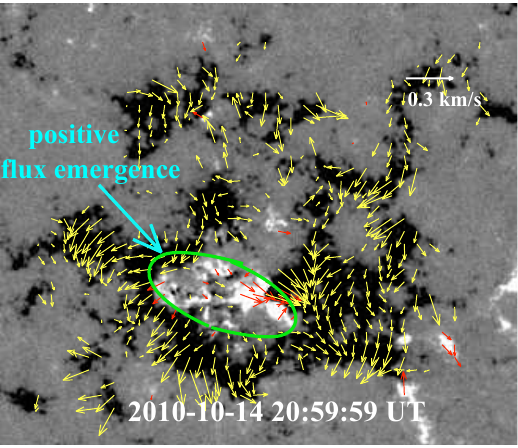}
}
\centerline{
\includegraphics[width=0.5\textwidth]{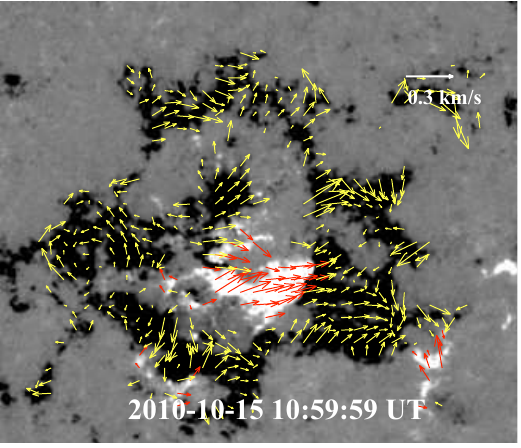}
\includegraphics[width=0.5\textwidth]{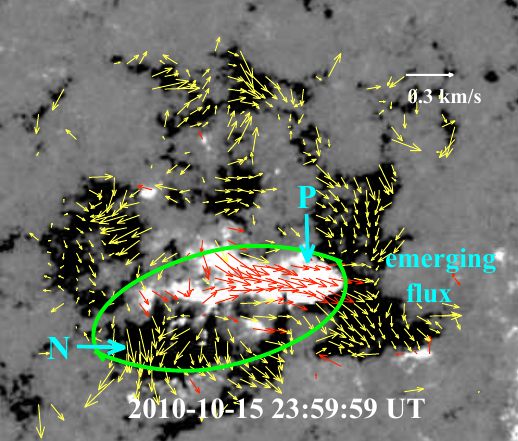}
}
\centerline{
\includegraphics[width=0.5\textwidth]{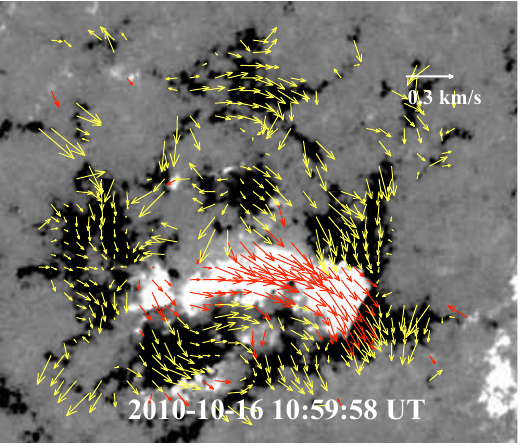}
\includegraphics[width=0.5\textwidth]{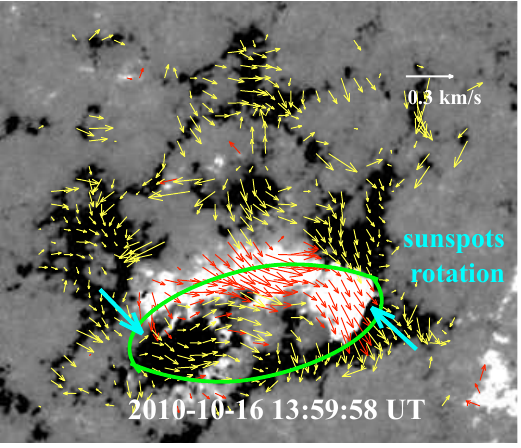}
}
\centerline{
\includegraphics[width=0.5\textwidth]{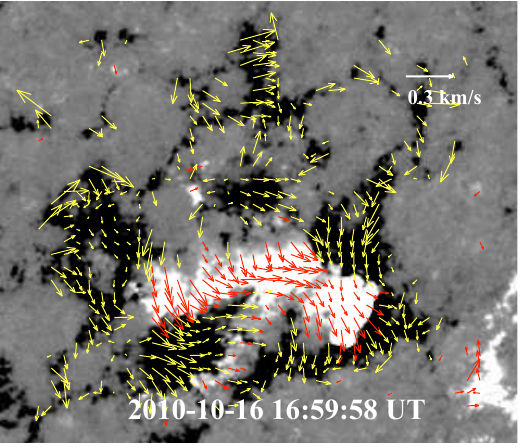}
\includegraphics[width=0.5\textwidth]{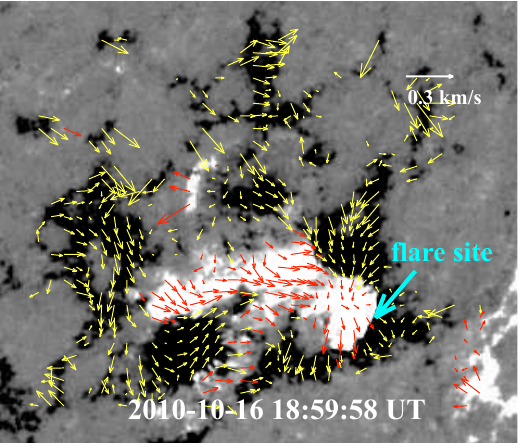}
}
\caption{Photospheric flow maps of the flare site obtained by DAVE technique using HMI magnetograms. The size of each 
image is $164^{\prime\prime}\times140^{\prime\prime}$. The magnetic field is scaled in the above images -200 G to +200 G.}
\label{flow}
\end{figure}
\begin{figure}
\centerline{
\includegraphics[width=0.8\textwidth]{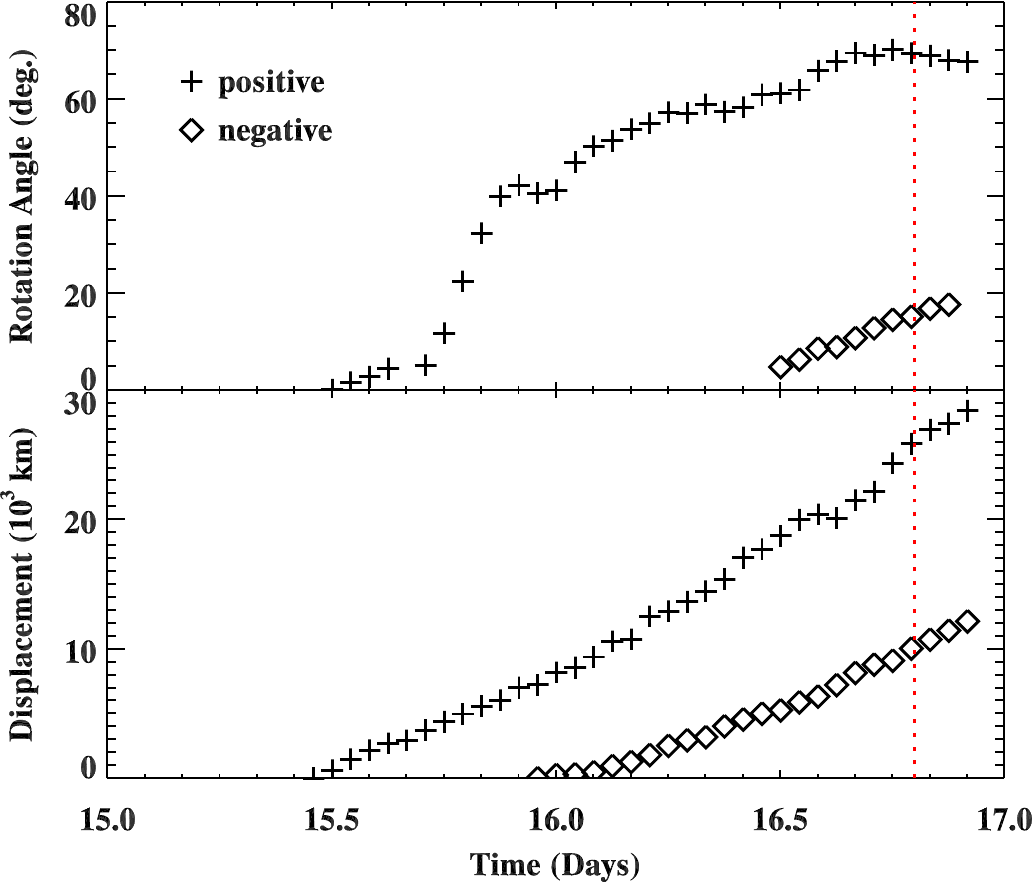}
}
\caption{Upper panel: Temporal evolution of the accumulated rotation angle of both the positive and negative polarity sunspots on 15 and 16 October 2010. The absolute values of the rotation angle have been plotted for the negative polarity sunspot. Lower panel: Accumulated displacement of the center position of both sunspots. The vertical dotted line indicates the flare peak time on 16 October 2010.}
\label{rot}
\end{figure}

\begin{figure}
\centerline{
\includegraphics[width=0.7\textwidth]{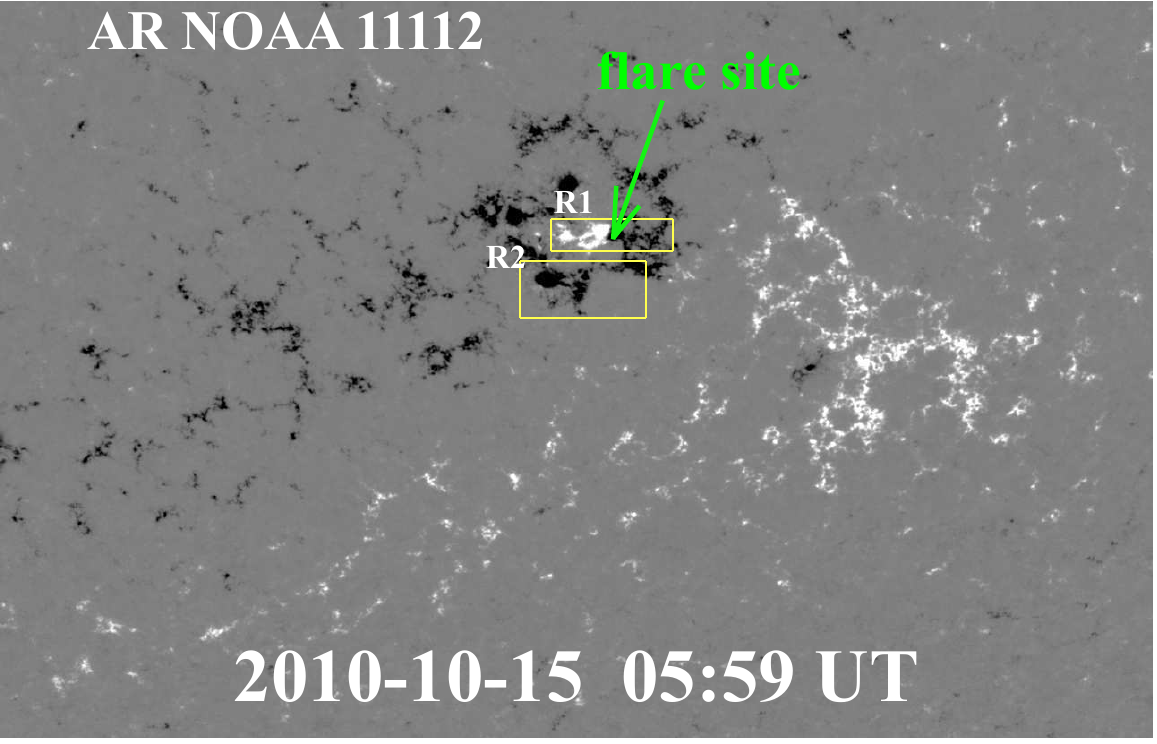}
}
\centerline{
\includegraphics[width=0.65\textwidth]{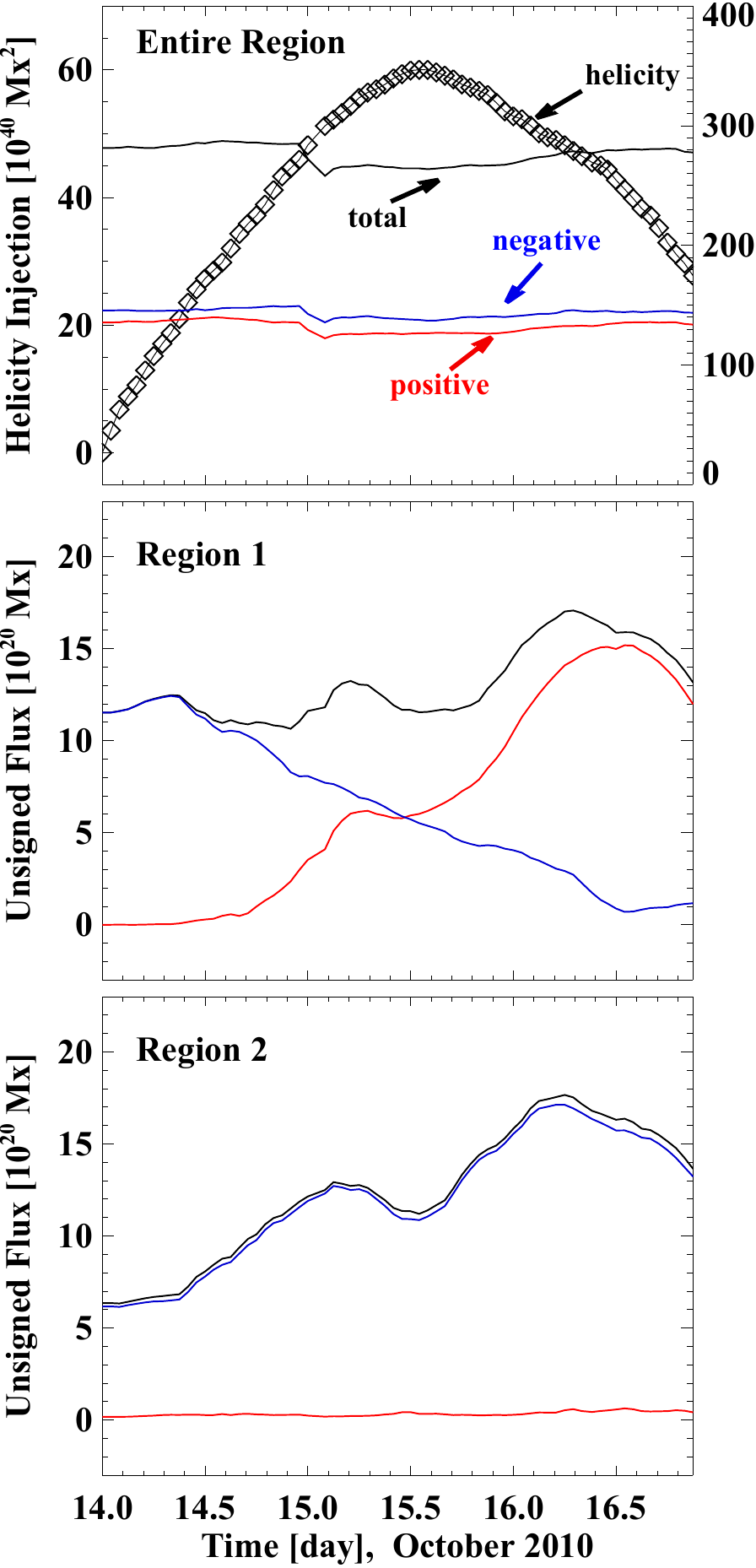}
}
\caption{Top panel: HMI magnetogram of the NOAA AR 11112 on 15 October 2010. Region 1 and region 2 are shown by boxes indicated by `R1' and `R2' respectively. The image size is $644^{\prime\prime}\times412^{\prime\prime}$. Bottom panel: Evolution of magnetic helicity (diamond), positive (red), negative (blue) and total (black) unsigned magnetic flux profiles beginning from 14 October 2010.}
\label{helicity}
\end{figure}
\section{Magnetic Field Evolution}
 We used the HMI magnetograms to follow the evolution of magnetic fields before and after the flare.  HMI is designed to study oscillations and the magnetic field at the photosphere \cite{scherrer2012}. It observes the full solar disk (4K$\times$4K) at 6173 \AA \ with a resolution of 1$^{\prime\prime}$ (0.5$^{\prime\prime}$ pixel$^{-1}$). The typical cadence for magnetograms is 45 s. The magnetogram image shows the photospheric magnetic field, with black and white indicating negative and positive polarities, respectively. We have used the HMI continuum images to follow the morphological evolution of the sunspots.
In order to see the morphological evolution of the sunspots at the photosphere and associated responses at the coronal heights, we compared co-temporal HMI continuum, magnetograms and AIA 94 \AA \ EUV images. Figure \ref{cont} displays the HMI continuum, magnetograms as well as AIA images from 14 October to 16 October 2010. The top row panels display pre-flux-emergence images. The emergence of opposite polarity sunspots started at 18:00 UT on 14 October 2010 and the sunspot pair grew continuously (please see HMI  movie). The panels in the second row from the top show the emergence of a bipole indicated inside the ellipse in HMI images. It is interesting to note that we see the emerging of ``anemone-type" coronal loop system in AIA 94 \AA \ image (indicated by arrow) at the same time. After the emergence,  positive polarity sunspot (`P') shows rotation in the clockwise direction. On the other hand, the negative polarity sunspot (`N') rotates in anticlockwise direction (detected in HMI images).  The comparison between  HMI and AIA images reveals that the footpoints of the emerged loop system are anchored in the emerged opposite polarity sunspots and moves as the sunspots move/rotate. This is a good example of emergence of a flux tube from below the photosphere to coronal heights. Further, the shape of the emerged loop seems to have a twisted field configuration. The structure/orientation of the field lines in the flux tube can be viewed. The helicity sign can be
characterised by the presence of elongated polarities, called ``magnetic tongues" of the emerging flux-rope (refer to Figures 1 and 3 in \inlinecite{luoni2011} for the detail way to determine the helicity sign). Therefore, here the line of sight magnetic field distribution in the emerging bipole indicates negative (left-handed) helicity.

\subsection{Photospheric Flow Maps}
 To deduce the photospheric flows, we selected hourly magnetograms from 14-16 October 2010 (before the flare event). We have used Differential Affine Velocity Estimator (DAVE) technique \cite{schuck2005,schuck2006,chae2008} for the estimation of photospheric flows. After many trials, we choose the width of Gaussian window 10$^{\prime\prime}$. Figure \ref{flow} displays selected flow maps  obtained with the DAVE technique. It is interesting to see the emergence of positive flux on 14 October 2010 and later the emergence of a bipole (indicated by `P' and `N') is more clear on 15 October 2010 (indicated by arrows). The positive flux emerges and moves towards the western side. The direction of flow arrows strongly suggests that the positive polarity sunspot rotates in the clockwise direction and pushes/slides towards the neighbouring opposite (negative) polarity field region in southwest direction. This helps in increasing the shear and build-up of magnetic energy in between opposite polarity field regions, which is being released later in the flare. Therefore, the interaction of the two opposite polarity regions (indicated as flare site) was the most likely cause for triggering the flare. On the other hand, the flow arrows for the emerging negative polarity sunspot (`N') shows anticlockwise rotation specially on 16 October 2010. 

\subsection{Measurement of Sunspot Rotation}
 The emerged opposite polarity sunspots showed translational as well as rotational motion before the flare occurrence. The emerged positive polarity has an elliptical shape. Its orientation can be described by an angle between its major axis and the equator in anticlockwise direction. The rotation rate is defined as the orientation change in two successive images of the sunspot. In order to measure the rotation angle of both sunspots, we use fit$\_$ellipse routine available in IDL library. The sunspot structure is defined as a region having magnetic field $\ge$10$\%$ of the peak field strength. This procedure provides the center position as well as the orientation of the best-fitted ellipse on the sunspot region. Figure \ref{rot} displays the temporal evolution of the accumulated rotation angle (upper panel) as well as accumulated displacement of the center position (lower panel) of positive (+) and negative (diamond) polarity sunspot. The emerged positive polarity sunspot started to rotate in the clockwise direction on 15 October 2010 at 12:00 UT and showed significant rotation ($\approx$40$^\circ$) till 23:00 UT. Later, it continued to rotate gradually on 16 October 2010 and the rotation angle becomes nearly constant 2-3 hours before the flare occurrence. Therefore the total rotation angle for the positive polarity sunspot was $\approx$70$^\circ$ within 30 hours. On the other hand, the emerged negative polarity sunspot does not show any detectable significant rotation on 15 October,  while it shows slow rotation in the anticlockwise direction on 16 October 2010 (see Figure \ref{rot} upper panel). The total rotation angle for the negative polarity sunspot was $\approx$20$^\circ$ within 10 hours. Therefore, the rotation rates for the positive and negative polarity sunspots are $\approx$2.3 and 2$^\circ$ h$^{-1}$ respectively, which are comparable. Apart from the rotational motion, both sunspots also show a translational motion (divergence) after the emergence (refer to lower panel of Figure \ref{rot}). The positive polarity sunspot shows a significant displacement after the emergence and moves $\approx$26 Mm within 30 hours in the southwest direction. On the other hand, negative polarity sunspot moved westward $\approx$12 Mm within 22 hours. The significant and continuing displacement as well as the rotation of positive polarity sunspot helped in the build-up of magnetic energy to trigger the flare. 

\subsection{Estimation of Helicity Injection}
 Magnetic helicity is a useful measure for global complexity and non-potentiality of a magnetic field system \cite{berger1984,pevtsov2008,dem2009} and it has been studied to understand an energy build-up process, pre-flare conditions, and triggering mechanisms of flares \cite{kusano2003,labonte2007,park2008,park2010}. We therefore determined magnetic helicity injection through the photospheric surface of the active region (AR) NOAA 11112 to quantify the temporal evolution of complexity built up by motions of magnetic fragments in sunspot areas of the flare-producing active region. 

To measure the magnetic helicity injection, we first calculated helicity flux density $G_{\theta}(\textit{\textbf{x}},t)$ (i.e., helicity injection per unit area per unit time) at a position $\textit{\textbf{x}}$ on the photospheric surface of AR 11112 and a specific time $t$ using the helicity flux density formula proposed by \inlinecite{pariat2005} and the numerical calculation method developed by \inlinecite{chae2007}: 
\begin{equation}
G_{\theta}(\textit{\textbf{x}},t)=-\frac{B_{n}}{2\pi} \int_{S'} \left(\frac{\textit{\textbf{x}}-\textit{\textbf{x}}\,'}{|\textit{\textbf{x}}-\textit{\textbf{x}}\,'|^{2}} \times [\textit{\textbf{u}}-\textit{\textbf{u}}'] \right)_{n} B_{n}' \,dS',
\end{equation}
where the subscript $n$ denotes the vertical component to the photospheric surface $S'$ of the active region. $B_{n}$ is the magnetic field component perpendicular to $S'$ and it is derived from the line-of-sight component $B_{l}$ by assuming that the transverse component of magnetic fields is negligible compared to $B_{l}$ (i.e., $B_{n}$ = $B_{l}$/$\cos{\psi}$ where $\psi$ is the heliocentric angle). $\textit{\textbf{u}}$ is the velocity of the apparent horizontal motion of field lines determined by applying the differential affine velocity estimator (DAVE) method \cite{schuck2006}. In this study, $G_{\theta}(\textit{\textbf{x}},t)$ was determined using SDO/HMI line-of-sight magnetogram data in the time span of 14 October 2010, 00:00 UT through 16 October 2010, 21:00 UT with a 1-hour cadence. Then the amount of helicity injection $H(t)$ through a local region $S_{R}$ in the active region is determined by integrating $G_{\theta}(\textit{\textbf{x}},t)$ with respect to the area of $S_{R}$ and time:
\begin{equation}
H(t)=\int_{t_{0}}^{t} \int_{S_{R}} G_{\theta}(\textit{\textbf{x}},t)\,dS_{R}\,dt,
\end{equation}
where $t_{0}$ is the start time of the magnetogram data set under investigation.

We have estimated the helicity injection, positive flux, unsigned negative and total unsigned magnetic flux in the entire active region as well as magnetic flux profiles in the regions of the localized emerged polarities. Top panel of Figure \ref{helicity} displays the HMI magnetogram of the AR, where R1 and R2 indicate the emerging flux regions for positive and negative polarities respectively. The flare site is indicated by an arrow.  The temporal variations of helicity injection (diamond), positive (red), unsigned negative (blue) and total (black) unsigned magnetic flux have been displayed in the bottom panel of Figure \ref{helicity}. The entire AR shows the injection of relatively more positive helicity starting from 14 October 2010, peaking ($\approx$60$\times$10$^{40}$ Mx$^2$) at 12:00 UT on 15 October 2010 and later it presents a relatively more negative helicity injection.
Region 1 (R1) shows the emergence of positive flux on 14 October ($\approx$12:00 UT), first peak on 15 October and enhances significantly ($\approx$15$\times$10$^{20}$ Mx) on 16 October before the flare triggering. Please note that the decrease in the negative flux in R1 is due to the pushing/sliding motion of the positive polarity, which make the negative polarity to move out from the box region R1. Region 2 (R2, bottom panel) displays the increase in the negative flux (maximum value$\approx$17$\times$10$^{20}$ Mx) same as the increase of the positive flux in R1. 
The entire active region has a magnetic inversion line, which is oriented nearly at 45$^{\circ}$ to the solar equator. As it was shown by \inlinecite{demoulin2002}, the helicity injection by differential rotation changes sign when the AR's inversion line becomes more tilted than 45$^{\circ}$. On the south hemisphere, the initial positive helicity injection in the entire region later becomes negative, i.e. the total coronal helicity stats to be depleted by differential rotation. We believe that this may be the case here, besides the negative helicity injection by the new bipole. The entire picture of magnetic helicity in the active region might be as follows: (1) at first, until the middle of 15 October, positive helicity was injected through the photospheric
surface of a much larger and dispersed bipole ({\it i.e.,} pre-existing bipole) which covers the
entire region of NOAA 11112. But negative helicity starts to be injected through the pre-existing
bipolar region by the differential rotation from the middle of 15 October, and (2) the newly
emerging bipole has pre-twisted magnetic fields with negative sign, which can be estimated from the study of \inlinecite{luoni2011} and their footpoints show the
rotation in opposite directions due to bodily rotation of the bipolar flux tube (rotating as a whole).
The important thing to note is that the emerging flux profiles for the positive and negative polarities are quite similar in region R1 and R2, which strongly suggests the emergence of a flux tube from below the photosphere. The coronal signature of emerging flux in the form of ``anemone-shaped" loop system, is confirmed in AIA 94 \AA \ images (Figure \ref{cont}). 
 

\begin{figure}
\centerline{
\includegraphics[width=0.6\textwidth]{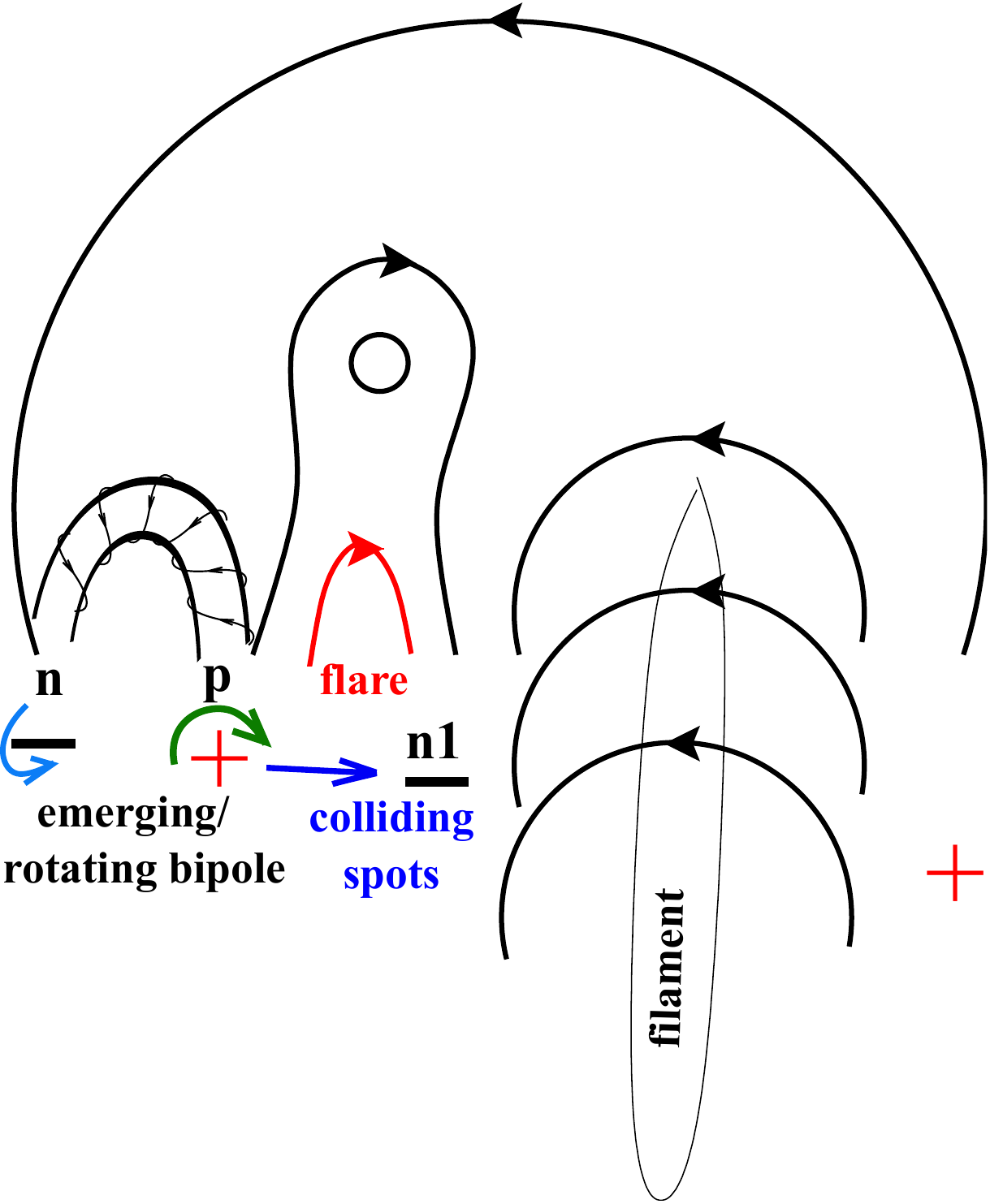}
}
\caption{Schematic cartoon showing the emerged flux tube, plasmoid ejection and associated magnetic reconnection at the flare site.}
\label{cart}
\end{figure}
\section{Discussion and Conclusions}
 
We have analysed the multiwavelength observations of the M2.9/1N flare occurred on 16 October 2010 from AR NOAA 11112. 
 SDO Observations reveal the emergence of flux tube within the existing AR from below the photosphere to the corona. The footpoints of the emerged bipole show the rotation in opposite directions. The positive polarity sunspot shows a significant rotation in the clockwise direction during $\approx$30 h before the flare, whereas negative polarity sunspot shows a relatively smaller rotation in the anticlockwise direction during $\approx$10 h before the flare triggering. 
Moving/rotating positive polarity sunspot continuously pushes the negative polarity field region towards southwest direction. The shear motion of this sunspot helped in the build-up of magnetic energy and the flare was triggered when the magnetic configuration became unstable. Moreover, the photospheric flow map also confirms the sliding motion of the positive sunspot in the southwest direction. \inlinecite{kurokawa1987} found two type of processes for the development of magnetic shear configurations between sunspots: (a) collision of two sunspots of opposite magnetic polarities, and (b) successive emergence of twisted magnetic flux ropes. He suggested that
the second process might be essential for the production of big flares. The present event indicates both processes, {\it i.e.}, emergence as well as the sunspot collision, which are responsible for the
development of magnetic shear to trigger the M-class flare.

 Figure \ref{cart} displays the schematic cartoon to
show the magnetic field environment of the flare site. The energy build-up and release process can be explained in the following way: (i) `n' and `p' displays the emerging and rotating bipole in opposite directions, (ii) field lines in the emerged loop shows  left-handed twist, whose footpoints are anchored in the rotating spots `n' and `p', (iii) clockwise rotating/sliding sunspot `p' collides with opposite polarity spot `n1', (iv) magnetic reconnection takes place in between colliding spots (`p' and `n1') and a plasmoid was ejected followed by an M-class flare. The first reconnection most likely started in the lower solar
atmosphere, which was confirmed by the
high frequency decimetric radio bursts profiles. The location of the RHESSI hard X-ray sources
and initial ribbon brightening during the flare impulsive phase suggest the
reconnection of field lines between two opposite polarity colliding spots.
 The plasmoid ejection during the flare impulsive phase is an indirect evidence 
of magnetic reconnection \cite{shibata1995}.

The emerging bipoles generally show the rotation in the same direction, which generates a sheared polarity inversion line. This process creates the sigmoid structure in the corona that shows `S'
or `inverse-S' shape in soft X-rays \cite{pevtsov1995,rust1996,canfield1999}. These sigmoid structures are generally considered to produce flares/CMEs. The present event shows the emergence of a flux tube associated with footpoints/sunspots rotation in opposite directions. This may explain why sigmoid was not formed and why the flux tube did not erupt. Although the footpoints rotate in opposite
directions, the total rotation-angles are different, which implies that the rope was twisted by this
process. A bodily rotation of a flux tube (rotating as a whole) injects
no net helicity in the system, though locally rotation is seen (refer to \opencite{pariat2007}). However, the observation that the two opposite polarity spots rotate in opposite directions indicates  that there is an additional global positive, right-handed rotation of the emerging flux rope (i.e. the flux rope is rotating as a whole). We believe that here we have a superposition of a flux rope emerging with negative helicity (as
shown by the magnetic tongue/tails structure, and a smaller rotation of the flux rope as a whole).
\inlinecite{yan2008} proposed that the two sunspots with
the same rotating direction have higher flare productivity than
those with opposite rotating directions, and this scenario
makes it easier to store magnetic energy and increase the helicity
of the flux tube. Indeed, the present AR was not large flare/CME productive and the first flare (M2.9) was observed on 16 October 2010 associated with a slow CME (speed$\approx$350 km s$^{-1}$).

Our observations show the flux emergence and rotation of sunspots simultaneously. This reveal the  important aspects about the origin 
and effect of sunspot rotation, which is consistent with the
idea suggested by many authors, that the rotation might result from the 
emergence magnetic fields pre-twisted below the photosphere 
\cite{brown2003,tian2006,min2009}. 
\inlinecite{chae2003} proposed that a variety of shearing motions
 and rotational motions observed in ARs may be driven by the torque
caused by the expansion of the coronal part of the twisted flux tube. 
\inlinecite{min2009} suggested that if an initially twisted flux tube emerges, the coronal segment of the tube is likely to
 expand due to the pressure difference between the interior and the corona. In this case, a
 magnetic-torque imbalance \cite{jockers1978} occurs between the expanded part (corona) and the unexpanded
 part (interior). So, the Lorentz force and the related torque on the photospheric boundary drive the sunspot rotation
 We also notice the expansion in the emerging flux tube in the corona, which is in agreement with the above explanation.

In conclusion, we presented the multiwavelength observations of an M-class
flare to show that its energy build-up was done by shearing and rotation of the positive polarity sunspot.
The flare was triggered by the collision of positive polarity sunspot with the surrounding opposite
polarity field region. Using the
high spatial and temporal data from space and ground based instruments, further
studies should be performed in order to shed more light on the flare triggering processes.  

\begin{acks}
We express our gratitude to the referee for providing the constructive comments/suggestions, which improved the manuscript considerably. SDO is a mission for NASA's Living With a Star (LWS) Program. We are thankful for the radio data obtained from RSTN network and GBRSBS. Global High Resolution H$\alpha$ Network is operated by the Space Weather Research Lab, New Jersey Institute of Technology. PK thanks to Prof. P.F. Chen and Dr. A. K. Srivastava for fruitful discussions. This work has been supported by the ``Development of Korea Space Weather Center" project of KASI, and the KASI basic research fund. 
 
\end{acks}
\bibliographystyle{spr-mp-sola}
\bibliography{reference}  
\end{article} 
\end{document}